\documentclass{kluwer}    
\newdisplay{guess}{Conjecture}

\usepackage{graphicx}
\input{tcilatex}

\begin{document}                    
\begin{article}

\begin{opening}         
\title{Relativistic versus Newtonian orbit model: \\
the Relativistic Motion Integrator (RMI) software. \\
Illustration with the {LISA} mission.} 
\author
{Sophie \surname{Pireaux}\footnote{{\tiny Department 1\newline
Observatoire Royal de Belgique, 3 avenue Cirulaire, 1180 BRUSSELS, BELGIUM.\newline
Tel: ++32(0)2 373 67 53 and Fax: ++32(0)2 374 98 22.\quad E-mail:
sophie.pireaux@oma.be\newline
Previously working in UMR 6162, ARTEMIS, OCA.}} 
Bertrand \surname{Chauvineau}\footnote{{\tiny UMR 6162, ARTEMIS,\newline
Observatoire de la C\^{o}te d'Azur (OCA), avenue de Copernic, 06130 GRASSE, FRANCE.}}
and Aurelien \surname{Hees}\footnote{{\tiny Department 1\newline
Observatoire Royal de Belgique, 3 avenue Cirulaire, 1180 BRUSSELS, BELGIUM.}}
}  
\runningauthor{S. Pireaux and B. Chauvineau and A. Hees}
\runningtitle{Relativistic versus Newtonian orbit model: the {RMI} software.}
\institute{Observatoire Royal de Belgique (ORB),\\
Observatoire de la C\^{o}te d'Azur (OCA)}
\date{25th June, 2009}

\begin{abstract}
The Relativistic Motion Integrator (RMI) consists in 
integrating numerically the EXACT relativistic equations of 
motion, for a given  metric (corresponding to a gravitational field at first post-Newtonian order or higher), 
instead of Newtonian equations plus relativistic corrections.\\
The aim of the present paper is to validate the method, and to illustrate how RMI 
can be used for space missions to produce relativistic ephemerides of test-bodies (or satellites). 
Indeed, nowadays, relativistic effects have to be taken into account, and comparing a RMI model 
with a classical keplerian one helps to quantify such effects. 

LISA is a relevant example to use RMI. 
This mission is an interferometer formed by three spacecraft which aims at
the detection of gravitational waves. 
A precise orbit model for the LISA spacecraft is needed not only
for the sake of satellite ephemerides but also to compute the photon flight time
in laser links between spacecraft, required in LISA data pre-processing in order to reach the
gravitational wave detection level.

Relativistic effects in LISA orbit model needed to be considered and quantified.\\ 
Using RMI, we show that the numerical classical model for LISA orbits in the gravitational field of 
a non-rotating spherical Sun without planets can be wrong, 
with respect to the numerical relativisitic version of the same model, 
by as much as about ten kilometers in radial distance 
during a year and up to about 60 kilometers 
in along track distance after a year...  
with consequences on estimated photon flight times.

We validated RMI numerical results (using a metric following the International Astronomical Union -IAU- 2000 resolutions) 
with an analytical developpement (up to first order
in eccentricity and up to first order in $GM/c^{2}$, where $G$ is Newton's
constant, $M$, the solar mass and $c$ the speed of light in vacuum).

Finally, the RMI relativistic numerical approach is soon more efficient than 
the analytical development. Moreover, RMI extends to other cases (planetocentric, instead of barycentric) and can be applied to other space missions.
\end{abstract}


\end{opening}   


\section{Introduction}

\qquad Due to considerable increase of the accuracy level in modern space missions
in the recent years, or expected in close-future missions, relativistic gravitational effects must be 
considered when computing spacecraft ephemerides.\\
Indeed, the Schwarzschild radius ($2GM/c^{2}$) of the Earth is of the order of one centimeter; while that of the Sun is of 
the order three kilometers. The first corresponds to the order of magnitude of the precision in current space geodesy; while the second, 
to the precision requested in some future space mission such as in LISA (2015). 
The relativistic Lense-Thirring effect has already been partially detected with LAGEOS Earth orbiting satellites \cite{Ciu2004}. 
Numerical integrations in the post-Newtonian approximation versus Newtonian ones have shown the relevance of relativistic effects 
in the orbit of the future GAIA mission (2011) \cite{Klio2005}, since the GAIA spacecraft must be controlled with an accuracy of 0.6mm/s. 
Owing to the above motivations, the present work is dedicated to 
a numerical relativistic model 
for a generic space mission. 
\newline

The method RMI (Relativistic Motion Integrator), a
fully consistent general relativistic approach \cite{Pi2005}, \cite{Pi2006a} consists in integrating
numerically the EXACT relativistic equations of motion for a given metric. 
The advantages of the method are the following. All relevant relativistic
effects are taken into account if a gravitational metric adapted to the precision of
measurements is chosen. The approach is relativistically consistent, 
and safer than adding relativistic corrections by hand to a computation first developed 
in a Newtonian framework. The RMI
approach \emph{natively} contains all the gravitational classical and relativistic effects at the
corresponding order of the metric, including all the couplings between these
effects at the corresponding order with respect to the metric chosen. 
This is a serious advantage over a Newtonian-plus-relativistic-corrections 
approach such as is implemented in
commonly used orbit determination softwares. These perturbation approaches become more and 
more questionable as the requested precision increases, requiring a larger number of 
relativistic effects to be taken into account. 
RMI could help to point out deficiencies in common softwares.\\
The standard approach to integrate the relativistic differential equations 
of motion are the Einstein-Infeld-Hoffmann (EIH) equations of motion (see \cite{Bru92},  
\cite{Bru04}, \cite{Bru07}, \cite{DSX91}, \cite{DSX92}, \cite{DSX93}, \cite{DSX94}, 
\cite{Moy2000} and references therein).
EIH equations are an analytical first order post-Newtonian (1PN) development 
of the exact relativistic equations of motion.
The advantage of the RMI method over the standard integration of EIH equations 
is that RMI straightaway numerically integrates 
the equations of motion for a chosen metric provided at a given PN order (whether 1PN or higher).
Hence, if, according to new IAU (or else) resolutions, a more appropriate/precise metric 
than the present Barycentric Coordinate Reference System (BCRS) metric for the motion 
in the Solar System or Geocentric Coordinate Reference System (GCRS) metric for 
planetocentric motion (with the Earth as central body) is recommended, RMI can 
straight away use that new metric... without the need to recalculate and implement 
new analytical developments. Only the metric module in the RMI software changes.
Indeed, separate modules in the RMI numerical method also
allow easy adaptations and updates for a given mission (number of plane
containing satellites, number of satellites per plane, initial conditions
-positions and velocities-), central body parameters (mass multipole
development of the gravitational potential, spin), planetary ephemerides,
IAU recommendations (metric, space-time transformations)... while keeping
the main body of the software unchanged.\newline

When wishing to illustrate how the RMI method can be used in space missions, 
LISA is a good candidate. However, the aim of the present paper is not to provide 
a thorough model of the LISA detector. Although some results obtained with RMI 
for LISA's orbit model are relevant for a LISA simulator.\\
The LISA (Laser Interferometer Space Antenna) mission \cite{LISA2000} is 
a space detector of gravitational waves in the $[\sim 10^{-4},\sim 10^{-1}]$ Hz
frequency band. Gravitational waves crossing the LISA quasi equilateral
triangular constellation are detected through the induced change in the
station inter-distances. The latter also depend on time, mainly due to the
gravitational field of the Sun \cite{Chau2005} around which LISA rotates, 20
degrees behind the Earth, and to that corresponding to planets; what we call
``geometry effects''.\newline
``Noise effects'' in LISA are orders of magnitude larger than
``gravitational wave source effects''. In order to reach the gravitational
wave detection level, a Time Delay Interferometry (TDI) method (see \cite{Dh2002}, 
\cite{Es2000}) must be applied to get rid of (most of) the laser
frequency noise and optical bench noise. The TDI method consists in
combining numerically data fluxes at the stations (rather than combining the
laser beams physically) with an appropriate delay. Hence, the so-called TDI
observables are symmetrized combinations of the different laser links with
appropriate delays (combination of photon-flight time between two stations
which correspond to station inter-distances) that cancel (almost all) the
laser frequency noise and optical bench noise. 
The TDI method is the crucial pre-processing of LISA data, 
before even trying a given strategy to detect any gravitational wave signal.\\
Therefore, in order to validate the new TDI technique and since a laboratory replica of the complex LISA mission is not totally achievable, the performance of  LISA TDI can only be studied with computer simulations of the different processes involved. Such is the aim of the LISACode software [\cite{Petiteau_et_al08}] developed by the LISAFrance group [\cite{LISA-France}], or of other simulators in the USA [\cite{Vallisneri05}, \cite{Cornish_et_al_04}].
Among the processes to be implemented in a LISA simulator, the orbit model of the spacecraft, providing positions, velocities and interdistances of spacecraft needed for TDI, is the subject of the present paper.

Relativistic effects in LISA needed to be considered and quantified. \\
In the framework of the LISA mission, in articles \cite{Chau2005} 
(see references therein for a generic approach) 
and \cite{Pi2006b}, the photon flight time problem, also sometimes referred
to as time transfer, and proper time scales of LISA spacecraft are tackled using
a consistent general \emph{relativistic} approach. However, the orbit model 
used to compute the initial positions and velocities of LISA spacecraft at
emission time needed in the time transfer simulation or in
proper-versus-coordinate time transformations is \emph{classical}. \\
And so is it still presently the case too in the TDI simulators named Synthetic LISA \cite{Vallisneri05}, LISA Simulator \cite{Cornish_et_al_04} and LISACode \cite{Petiteau_et_al08}.\\
In the preliminary optimal orbit design for LISA used by Hughes \cite{Hu2005},
LISA's orbit model is also purely classical (in presence of a
spherical non-rotating Sun with planets). The author looks for the optimal
set of orbital inclinations, eccentricities, semi-major axis, longitude of
the ascending nodes, arguments of perigee and initial mean anomalies 
$(a_{k},e_{k},i_{k},\Omega _{k},\omega _{k},M_{k0})$ of LISA spacecraft 
$(k=1,2,3)$ in order to minimize LISA's arm flexing according to certain
optimization criteria.

In the present article, we use RMI (assuming no non-gravitational forces for LISA spacecraft motion) 
to quantify the errors implied 
when a classical orbit model is adopted for LISA instead of a general
relativistic one for the same initial conditions (Barycentric Coordinate
Reference System -BCRS- positions and velocities of spacecraft). 
We first investigate the case of a classical \emph{circular} orbit of reference around a
spherical non-rotating Sun without planets, which we call the \emph{circular spherical symmetric case}. 
We then extend to \emph{eccentric} orbits and name this case the 
\emph{eccentric spherical symmetric case} (more specifically for LISA, $e\simeq0.0096$).\\
Our numerical estimate of relativistic versus Keplerian
orbit model for LISA with $e=0.0096$ shows that the difference between predicted barycentric
relativistic and classical radial distance reaches up to about $8-9$ km during a one-year mission 
and that the along track difference in orbits is about $54-59$ km after one year (i.e. after one classical period), according to the spacecraft considered, in the eccentric spherical symmetric case.\\
The relativistic versus classical modelling of LISA's orbit has repercussions on the flexing of LISA interferometric arms, 
the so-called breathing of the LISA constellation around its nominal arm-length value $L=5\cdot 10^{9}$ m. 
We show that a relativistic 
orbit model 
is relevant when studying photon time transfer needed in the TDI method; more specifically because the zeroth order is but the spacecraft inter-distance divided by the speed of light.\\

Since LISA eccentricity is small and because TDI and classical orbit models for LISA used by the Mock LISA Data Challenge (MLDC) \cite{Arnaud_et_al07} task force have been developped using first-order in eccentricity approximations, 
we provide a relativistic analytical check: 
a development up to first order in eccentricity and up to first order in $GM/c^{2}$, where 
$G$ is Newton's constant, $M$, the solar mass and $c$ the speed of light in vacuum) 
circular or eccentric spherical symmetric cases. \\
For the circular spherical symmetric case, the analytical development up to first order in $e$ and $GM/c^{2}$ (equations (\ref{analytical_check_einull_BCRS_orbit_tetha_solution}) and (\ref{analytical_check_einull_BCRS_orbit_radius_solution}) ) leads to small residuals (about $1$ cm in x-y-positions or along track distance and a few millimeters in radius) with respect to the RMI numerical relativistic model for LISA. However, in the eccentric spherical symmetric case, even for a small eccentricity such as LISA's ($e\simeq 0.0096$), the corresponding residuals are non negligible (reaching up to about $85$ m in along track distance) due to the $e^{2}$ and higher terms neglected 
in the analytical development; whereas RMI implicitly contains all order in $e$. Hence, the analytical development is soon surpassed by the 
numerical relativistic approach of RMI. This remark is even more relevant to space missions with important eccentricities.\\
The RMI method was furthermore validated in reference \cite{HeesPireaux09} (for BepiColombo or MarsNext mission) using a full 1PN development.
\\

The present paper is organized as follows. In Section 2, we recall the classical orbit model for LISA around a spherical symmetric Sun, which is to be our trial example for RMI along this paper.\\
In Section 3, we summarize the RMI relativistic numerical method and apply it to LISA with the appropriate initial conditions corresponding to the classical model. The numerical results obtained for LISA are then discussed. \\
In Section 4, we provide an analytical developpement (up to first order in eccentricity and in $GM/c^2$) to check RMI. \\ 
Finally, in Section 5, we conclude on the relevance of the RMI approach and on the main results obtained for LISA. \\
The annex \ref{numerical_integration_precision} discusses the numerical accuracy of the RMI method for LISA.

We adopt Einstein's summation convention on repeated
indices. Latin indices are for space coordinates, such as $l=1,2,3$; 
while Greek indices are for space-time coordinates, such as $\alpha = 0,1,2,3$ with \\
$x^{\alpha =0,1,2,3}=(c\cdot t,x,y,z)$. 
\\

\section{LISA classical orbit model in the spherical symmetric case}

\label{classical_orbitography}

\qquad Presently, within simulators testing LISA TDI (in the framework of the LMDC \cite{Arnaud_et_al07}), the following simplifications relative to LISA orbits are assumed. Each spacecraft follows perfectly a free-falling test mass that is itself perfectly shielded from non-gravitational forces and feels no constraints (for simplicity, one test-mass per spacecraft is modeled). 
As the gravitational field is concerned, solely a spherical non-rotating Sun is considered. 
The orbit model is classical. \\
In present LISA simulators for TDI, departures from the above assumptions on orbits are presently considered as part of the noise budget in TDI: among residual laser frequency and optical bench noises, scattered-light noise, detector shot noise, laser-beam pointing instability, acceleration noise, inertial noise and others (as specified in Table 1 of reference \cite{Petiteau_et_al08}).

For such a classical orbit model for the three LISA\ spacecraft 
$k=1,2,3$, in the BCRS, as in \cite{Dh2005}, the barycentric
coordinates $\left( x_{k},y_{k},z_{k}\right) $, for arbitrary initial
conditions, can be rewritten in terms of rotated Keplerian ellipses $\left(
x_{ell\ k},y_{ell\ k},z_{ell\ k}\right) $ with eccentricity $e\simeq0.0096$
as 
\begin{eqnarray}
\left( 
\begin{array}{l}
x_{k} \\ 
y_{k} \\ 
z_{k}
\end{array}
\right) &=&\Re ^{-1}\left( 
\begin{array}{l}
x_{ell\ k} \\ 
y_{ell\ k} \\ 
z_{ell\ k}
\end{array}
\right)  \label{classical_BCRS_orbit_coordinate_equations} \\
&&  \nonumber
\end{eqnarray}
\noindent with\ \medskip \newline
$\left( 
\begin{array}{l}
x_{ell\ k} \\ 
y_{ell\ k} \\ 
z_{ell\ k}
\end{array}
\right) \equiv \left( 
\begin{array}{l}
a\left( \cos \Psi _{k}-e\right)  \\ 
a\sqrt{1-e^{2}}\sin \Psi _{k} \\ 
0
\end{array}
\right) $\medskip \newline
\noindent
where $a$, $e$, $i$ and $\omega $ are the common semi-major axis, eccentricity, inclination and
argument of the periaster of the three spacecraft orbits, respectively; and \medskip \newline
$\Re ^{-1}\equiv \left( 
\begin{array}{lll}
\wp _{1} & \wp _{2} & \wp _{3}
\end{array}
\right) \ ,\smallskip $\newline
\noindent 
where the columns of the inverse rotation matrix are given by\newline 
\noindent
$
\begin{array}[t]{l}
\wp _{1}\equiv \left( 
\begin{array}{l}
+\cos \Omega _{k}\cos \omega -\sin \Omega _{k}\sin \omega \cos i \\ 
+\sin \Omega _{k}\cos \omega +\cos \Omega _{k}\sin \omega \cos i \\ 
+\sin \omega \sin i
\end{array}
\right) \ ,\smallskip  \\ 
\wp _{2}\equiv \left( 
\begin{array}{l}
-\cos \Omega _{k}\sin \omega -\sin \Omega _{k}\cos \omega \cos i \\ 
-\sin \Omega _{k}\sin \omega +\cos \Omega _{k}\cos \omega \cos i \\ 
+\cos \omega \sin i
\end{array}
\right) \ ,\smallskip  \\ 
\wp _{3}\equiv \left( 
\begin{array}{l}
+\sin \Omega _{k}\sin i \\ 
-\cos \Omega _{k}\sin i \\ 
+\cos i
\end{array}
\right) \ .\smallskip 
\end{array}
$
\newline
\noindent
Indeed, we start from a slightly different hypothesis with respect to Hughes' \cite{Hu2005}. 
We take common $(a,e,i)$ for the three spacecraft with optimal $e$, $i$ 
in order to minimize LISA's arm flexing in agreement with reference \cite{Na2005}:\\ 
$
\begin{array}[t]{l}
a=1\ \text{A.U}, \\ 
e=\sqrt{1+\frac{4}{\sqrt{3}}\frac{L}{2a}\cos \nu +\frac{4}{3}\left( \frac{L}
{2a}\right) ^{2}}-1, \\ 
i=arctg\left(\frac{\frac{L}{2a}\sin \nu }{\sqrt{3}/2+\frac{L}{2a}\cos \nu }\right)
\end{array}
$\medskip \newline
\noindent
where $\nu =\frac{\pi}{3}+\frac{5}{8}\frac{L}{2a}$ is the optimal inclination 
of the LISA triangle on the ecliptic and $L=5\cdot 10^{9}$ m is the average
interferometric arm-length. The longitude of the ascending node, $\Omega _{k}$, 
is particular to a given spacecraft $k$ and is given in terms of that of
the first one with a phase shift $\vartheta _{k}$: 
\[
\Omega _{k}=\Omega _{1}-\vartheta _{k}\quad \text{with }\vartheta _{k}\equiv
-2\left( k-1\right) \frac{\pi }{3}. 
\]
\noindent
The time parametrization of the orbits is given by the equation of the
eccentric anomaly $\Psi _{k}$ of each spacecraft, 
\begin{equation}
\Psi _{k}-e\sin \Psi _{k}=M_{k}\text{ ,}
\label{classical_eccentric_anomaly_equation}
\end{equation}
\noindent
with the mean anomaly 
\[
M_{k}=\frac{2\pi }{T}\left( t-t_{0}\right) +M_{k0} 
\]
\noindent
in terms of the orbital period, $T$, and the mean anomaly of spacecraft $k$ at
initial time $t_{0}$, that is $M_{k0}\equiv M_{k}(t=t_{0})$.\newline
Mean anomalies are related to that of the first spacecraft through the phase
shift: 
\[
M_{k}=M_{1}+\vartheta _{k}\text{ .} 
\]

BCRS position and eccentric anomaly equations used in \cite{Chau2005}
correspond to particular initial conditions ($ t_{0}=0,\ \omega =3\pi
/2,\ $ $\Omega _{1}=3\pi /2,\ M_{10}=0$) without any planets (which
means that both the initial time, $t_{0}$, and the initial mean anomaly of the first spacecraft, $M_{10}$, are completely arbitrary in that case).\\
We also recall that the time when spacecraft $k$ is at perihelion is given by 
\[
t_{kp}=t_{0}-\frac{M_{k0}}{n} 
\]
\noindent
with the mean
motion $n\equiv 2\pi /T=\sqrt{GM/a^{3}}$ from Kepler's 3rd law.

\section{Numerical \emph{native} relativistic orbit model}


\subsection{Exact relativistic equations of motion}

In General Relativity, the motion of a spacecraft is described by the
relativistic equation of motion, 

\begin{equation}
\frac{d^{2}x^{\alpha }}{d\tau ^{2}}=-\Gamma _{\beta \gamma }^{\alpha
}\cdot \frac{dx^{\beta }}{d\tau }\cdot \frac{dx^{\gamma }}{d\tau }
+K_{\beta }\left[ g^{\alpha \beta }-\frac{dx^{\alpha }}{d\tau }\cdot 
\frac{dx^{\beta }}{d\tau }\right]
\label{relativistic_BCRS_orbit_equations}
\end{equation}
\noindent
where $K_{\beta }$ is a quadri-``force'' encoding non-gravitational
forces; $\tau $, the proper time aboard the considered
spacecraft; and $\Gamma _{\beta \gamma }^{\alpha }$, Christoffel symbols
with respect to the metric. 
The relation between covariant and contravariant metric components being 
\begin{equation}
g^{\alpha \beta }\cdot g_{\beta \gamma}= \delta^{\alpha}_{\gamma} \label{cov_contra_metric_components}
 . 
\end{equation}
The four equations in (\ref{relativistic_BCRS_orbit_equations}) are redundant 
because of the normalization of the quadrivelocity.

In the case of LISA, assuming only one shielded test-mass per satellite, each
satellite follows a geodesic motion, that is $K_{\beta }=0$. 
Combining equations in (\ref{relativistic_BCRS_orbit_equations}), we can remove the proper 
time variable to rewrite the set of relativistic equations as 
\begin{equation}
\frac{d^{2}x^{l }}{dt^{2}}=\left[ -\Gamma _{\beta \gamma }^{l
}+\frac{1}{c}\Gamma _{\beta \gamma }^{0}\cdot \frac{dx^{l}}{dt}\right]
 \cdot \frac{dx^{\beta }}{dt}\cdot \frac{dx^{\gamma }}{dt}
\label{non_afine_geodesic_relativistic_BCRS_orbit_equations}
\end{equation}
\noindent

\subsection{Relativistic Motion Integrator (RMI) method applied to LISA}

\label{RMI_method}

The Relativistic Motion Integrator (RMI) method \cite{Pi2005}, \cite{Pi2006a}, 
consists in integrating numerically the \emph{exact} relativistic equations of
motion (\ref{relativistic_BCRS_orbit_equations}) for a given metric. \\
The numerical accuracy and stability of the RMI method for the LISA mission is validated in Annex \ref{numerical_integration_precision}.

When using the RMI method for LISA, rotating around the Sun, the appropriate
metric is the BCRS metric recommended by the IAU, International Astronomical Union, 2000
resolutions (see \cite{So2003} and references therein) 
and the corresponding isotropic coordinates.
The BCRS IAU 2000 metric neglects only terms at order $1/c^{5}$ and
above in $g^{00}$ or $g^{0l}$; and at order $1/c^{4}$ and above in $g^{lm}$. 
The IAU 2000 resolutions have been adopted in 2000 so to take into account the best precision of 
present and next future space experiments. That is experiments involving 
(or which can be translated in terms of) clocks, with accuracies better than a few parts in $10^{17}$ 
in fractional frequency and stabitities better than about $\sigma_{y}(\tau)=1\cdot 10^{-14}\tau^{-1/2}$ 
(Allan standard deviation), located at distances as close as $0.25$ A.U. from the Sun \cite{So2003}.\\
Note that most NASA and ESA space missions are modeled according to the EIH equations and corresponding 
relativistic algorithms described e.g. in \cite{Moy2000}. Unfortunately, reference \cite{Moy2000} was 
published around October 2000 and thus does not take into account the latest IAU2000 resolutions, 
published later. \cite{Moy2000} refers to IERS 1997 resolutions at the latest. 


\subsection{LISA initial conditions}

\label{initial_conditions}
\qquad 
We shall use the subscript $*_{cl}$ for classical quantities and $*_{rel}$ 
for the relativistic ones.\\
In our problem of comparing relativistic and classical LISA
ephemerides ($E$), we chose to take the same initial conditions ($IC$) in terms of
coordinate positions and velocities of spacecraft $k=1,2,3$ for both the relativistic
and the classical orbits. Indeed, we could have chosen to speak in terms of
same energy and momentum, but this does not reflect the way the actual space
mission will be planned and this does not easily provide insight in terms of what
is the error in predicted position and velocities. 
Hence, initial conditions of the relativistic model will be those BCRS 
$(x_{k},y_{k},z_{k};dx_{k}/dt,dy_{k}/dt,dz_{k}/dt)$ obtained by setting $t=t_{0}$
in the classical equations (\ref{classical_BCRS_orbit_coordinate_equations}) and 
(\ref{classical_eccentric_anomaly_equation}). \\
Note that this choice is not restrictive since, if the classical and relativistic $IC$ differ, \\
$
\begin{array}{ll}
E_{rel}(IC_{rel})-E_{cl}(IC_{cl})= & [E_{rel}(IC_{rel})-E_{cl}(IC_{rel})] \\ 
& +[E_{cl}(IC_{rel})-E_{cl}(IC_{cl})]
\end{array}
$\\
\noindent
and the second r.h.s. term in the above equation, not discussed in this paper, is but a classical problem.\\
The eccentricity of the numerical ephemerides for the eccentric case is that corresponding
to LISA spacecraft, $e\simeq0.0096$.
In our numerical simulation, we arbitrarily further chose 
$t_{0}=0$,$\ \omega =3\pi /2,\ \Omega _{1}=3\pi /2$
and$\ M_{10}=0$ in agreement with the initial conditions of paper \cite{Chau2005}.\\
Let us point out that this analysis could have been applied to Hughes' initial conditions \cite{Hu2005}. 


\subsection{Discussing numerical results for LISA in the spherical symmetric case}


The spherical symmetric model for LISA corresponds to a classical orbit of reference around a spherical non-rotating Sun without planets. Owing to this symmetry, the value of the inclination $i$ is irrelevant in order to compare relativistic versus classical ephemerides generated for LISA. 
Hence, we used the classical method without planets, described in Section \ref{classical_orbitography}, with $i$ set to $0$ and $e$ set to either $0$ (circular case) or $0.0096$ (eccentric case), to produce a numerical classical ephemeris for LISA $(x_{k},y_{k},z_{k},dx_{k}/dt,dy_{k}/dt,dz_{k}/dt)_{t}$. We then used the RMI method, described in the above Paragraphs \ref{RMI_method} and \ref{initial_conditions} with identical initial conditions, to produce a corresponding relativistic numerical ephemeris. We then used those two ephemerides, recorded as a function of BCRS time, to plot (relativistic - classical) quantities as a function of BCRS time every day during 365 days ($\simeq T=2 \pi / n $) such as in Section \ref{figures}. 


\subsubsection{Circular classical reference orbit case:}


\label{numerical_results_einull} 

\begin{figure}[h]
\begin{center}
\includegraphics[width=0.5\textwidth]{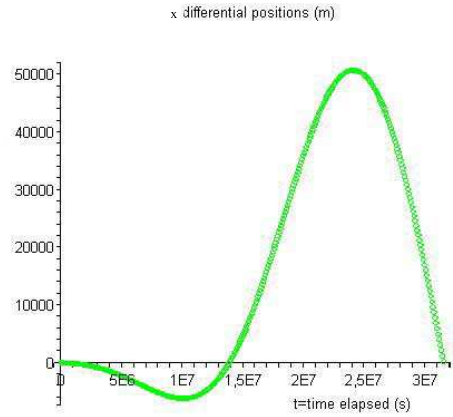}
\end{center}
\caption{Difference between numerical relativistic and classical position ephemerides for the LISA mission in the \emph{circular} spherical symmetric case:
 $x$ barycentric coordinate ($\delta x$).}
\label{numeric_diff_relat_kepler_X}
\end{figure}

\begin{figure}[h]
\begin{center}
\includegraphics[width=0.5\textwidth]{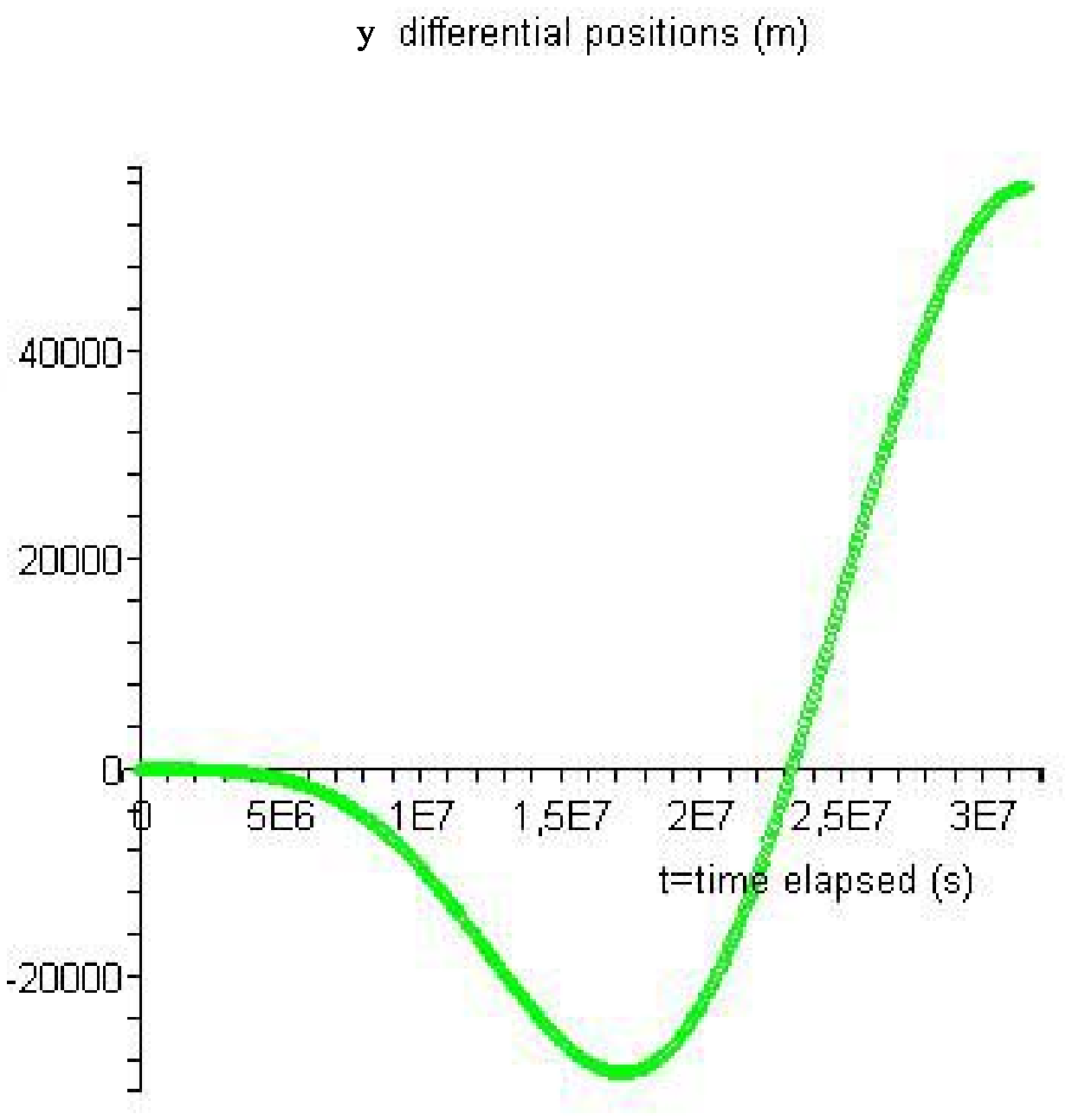}
\end{center}
\caption{Difference between numerical relativistic and classical position ephemerides for the LISA mission in the \emph{circular} spherical symmetric case:
 $y$ barycentric coordinate ($\delta y$).}
\label{numeric_diff_relat_kepler_Y}
\end{figure}

\begin{figure}[t]
\begin{center}
\includegraphics[width=0.5\textwidth]{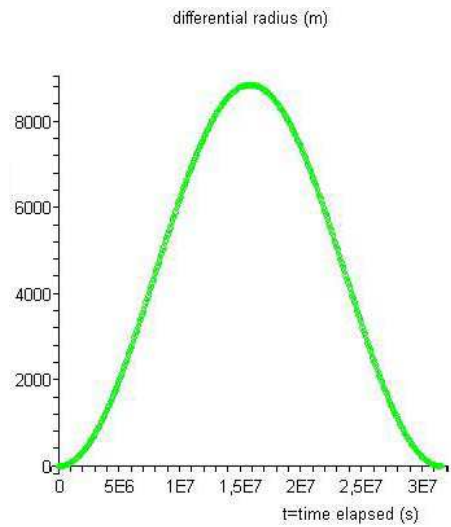}
\end{center}
\caption{Difference between numerical relativistic and classical position ephemerides for the LISA mission in the \emph{circular} spherical symmetric case:
 radial barycentric distance ($\delta r$).}
\label{numeric_diff_relat_kepler_radius}
\end{figure}

\begin{figure}[t]
\begin{center}
\includegraphics[width=0.5\textwidth]{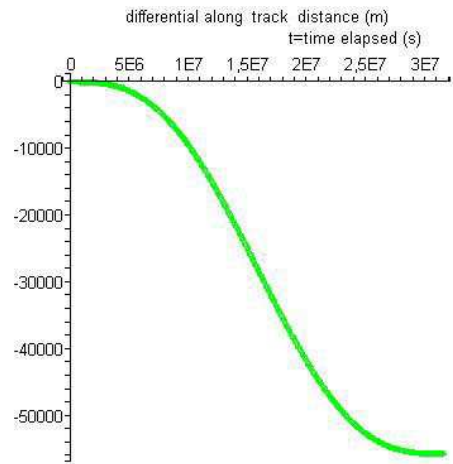}
\end{center}
\caption{Difference between numerical relativistic and classical position ephemerides for the LISA mission in the \emph{circular} spherical symmetric case:
 along track distance 
($\delta l\equiv r_{rel}\cdot \delta \theta \simeq a\cdot \delta \theta $).}
\label{numeric_diff_relat_kepler_along_track_distance}
\end{figure}

\begin{figure}[t]
\begin{center}
\includegraphics[width=0.5\textwidth]{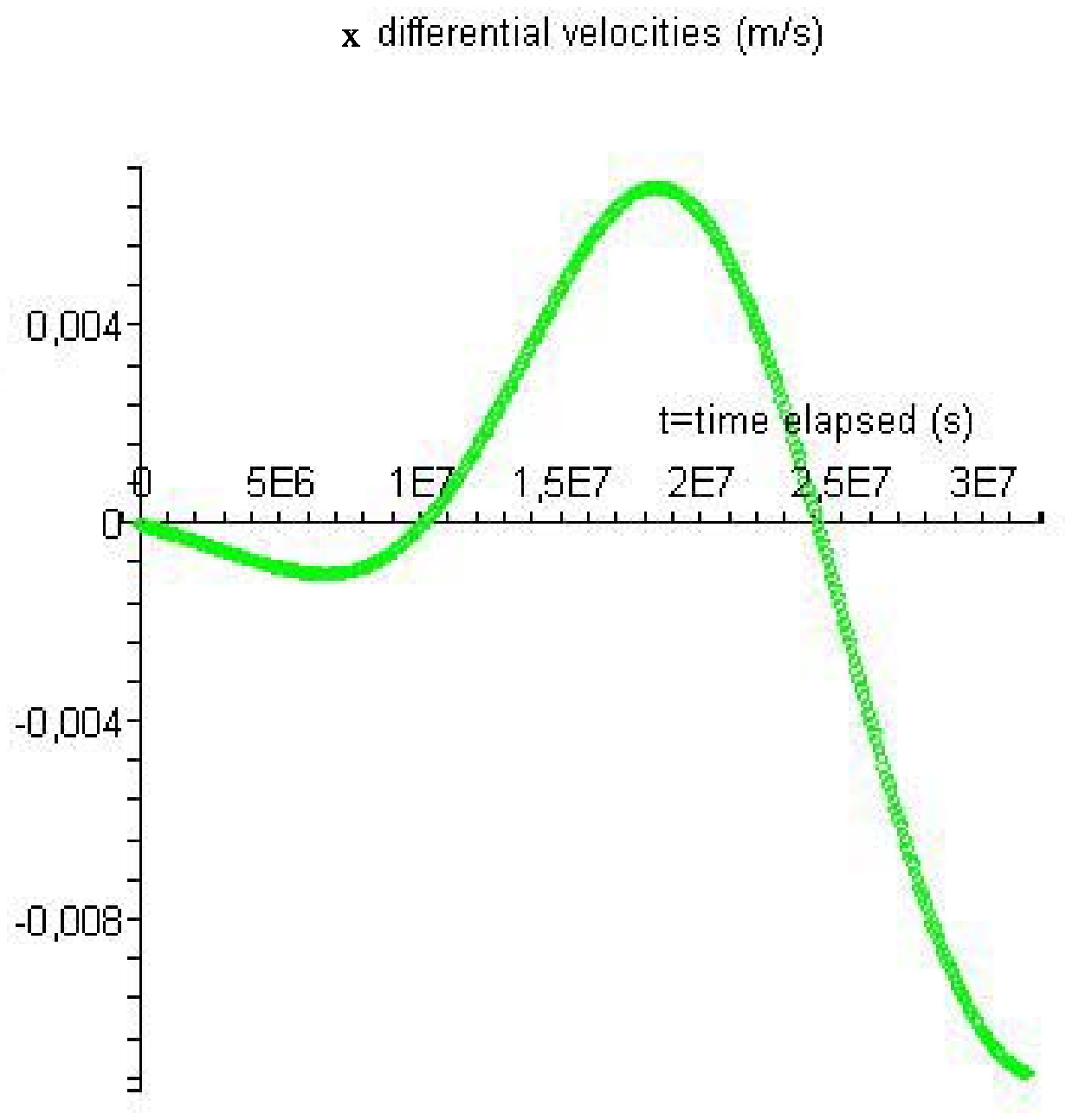}
\end{center}
\caption{Difference between numerical relativistic and classical velocity ephemerides for the LISA mission in the \emph{circular} spherical symmetric case:
 velocity component along the x barycentric coordinate axis ($\delta \stackrel{\bullet }{x}$).}
\label{numeric_diff_relat_kepler_X_dot}
\end{figure}

\begin{figure}[t]
\begin{center}
\includegraphics[width=0.5\textwidth]{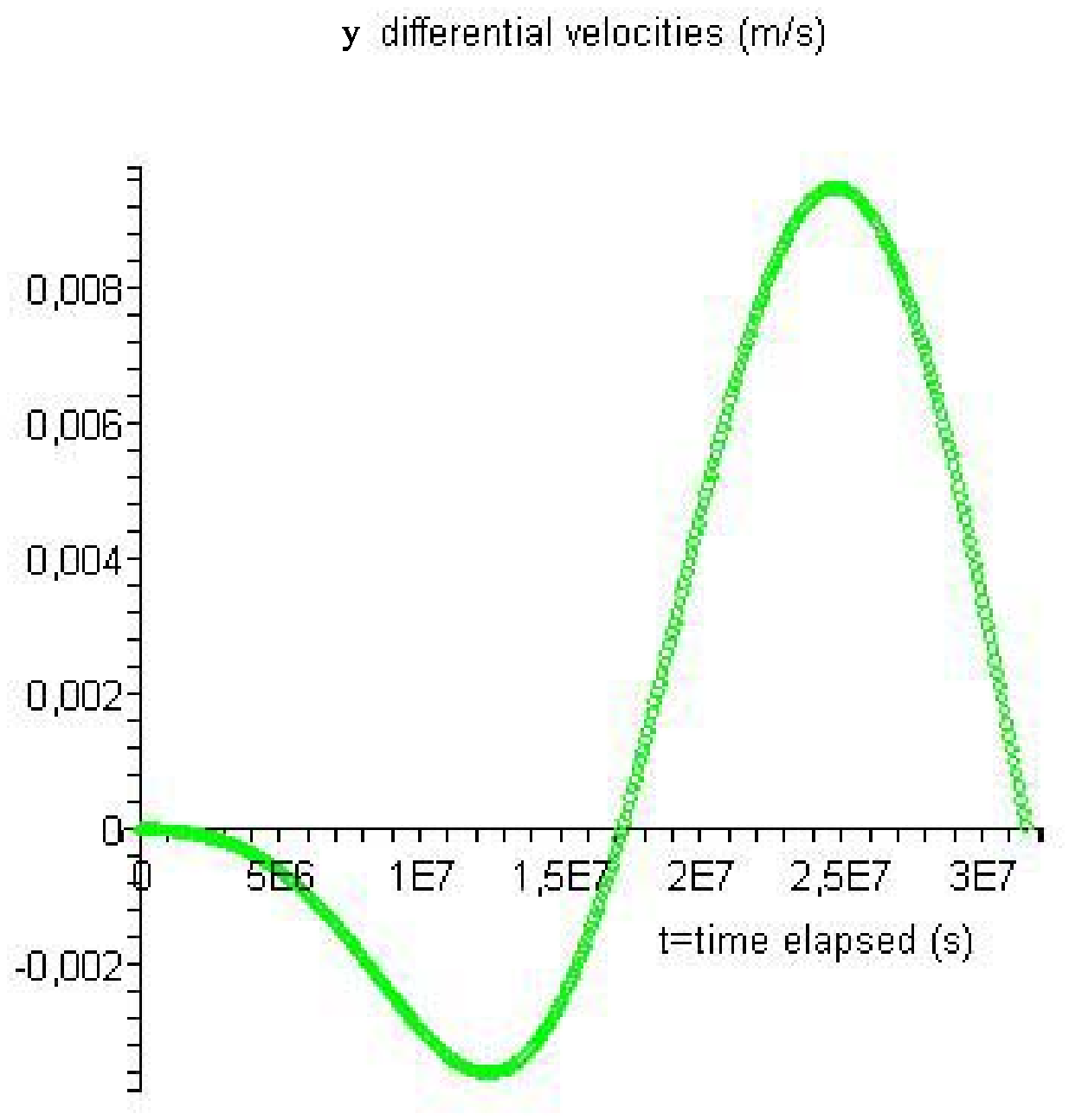}
\end{center}
\caption{Difference between numerical relativistic and classical velocity ephemerides for the LISA mission in the \emph{circular} spherical symmetric case:
 velocity component along the y barycentric coordinate axis ($\delta \stackrel{\bullet }{y}$).}
\label{numeric_diff_relat_kepler_Y_dot}
\end{figure}

From Figures \ref{numeric_diff_relat_kepler_X} and \ref{numeric_diff_relat_kepler_Y}, we found that the difference between predicted barycentric relativistic and classical x-y-positions reaches up to a maximum of about $51-56$ km during a one-year mission.\\
When speaking in terms of a difference in radial or along track distance between numerical relativistic and classical orbits, the above cited results translate into Figures \ref{numeric_diff_relat_kepler_radius} and \ref{numeric_diff_relat_kepler_along_track_distance}, respectively.  
We computed that the maximum difference in radius is about $8.9$ km while the along track difference in orbits after one classical period is about $56$ km for this circular spherical symmetric case. \\
The spacecraft is ahead on the classical orbit with respect to the relativistic one. \\
We see from Figure \ref{numeric_diff_relat_kepler_radius} that, having adopted a circular classical orbit of reference, the corresponding relativistic orbit is non-circular. \\
The difference in velocity components along the x- or y-BCRS axis as a function of time obtained are given in Figures \ref{numeric_diff_relat_kepler_X_dot} and \ref{numeric_diff_relat_kepler_Y_dot}, respectively. The difference between predicted barycentric relativistic and classical x-y-velocities reaches up to a maximum of about $0.007-0.010$ m/s during a one-year mission.
This agrees with the order of magnitude for the difference in position over one year. 


\subsubsection{Eccentric ($e=0.0096$) classical reference orbit case:}

\label{numerical_results_inull} 

\begin{figure}[h]
\begin{center}
\includegraphics[width=0.5\textwidth]{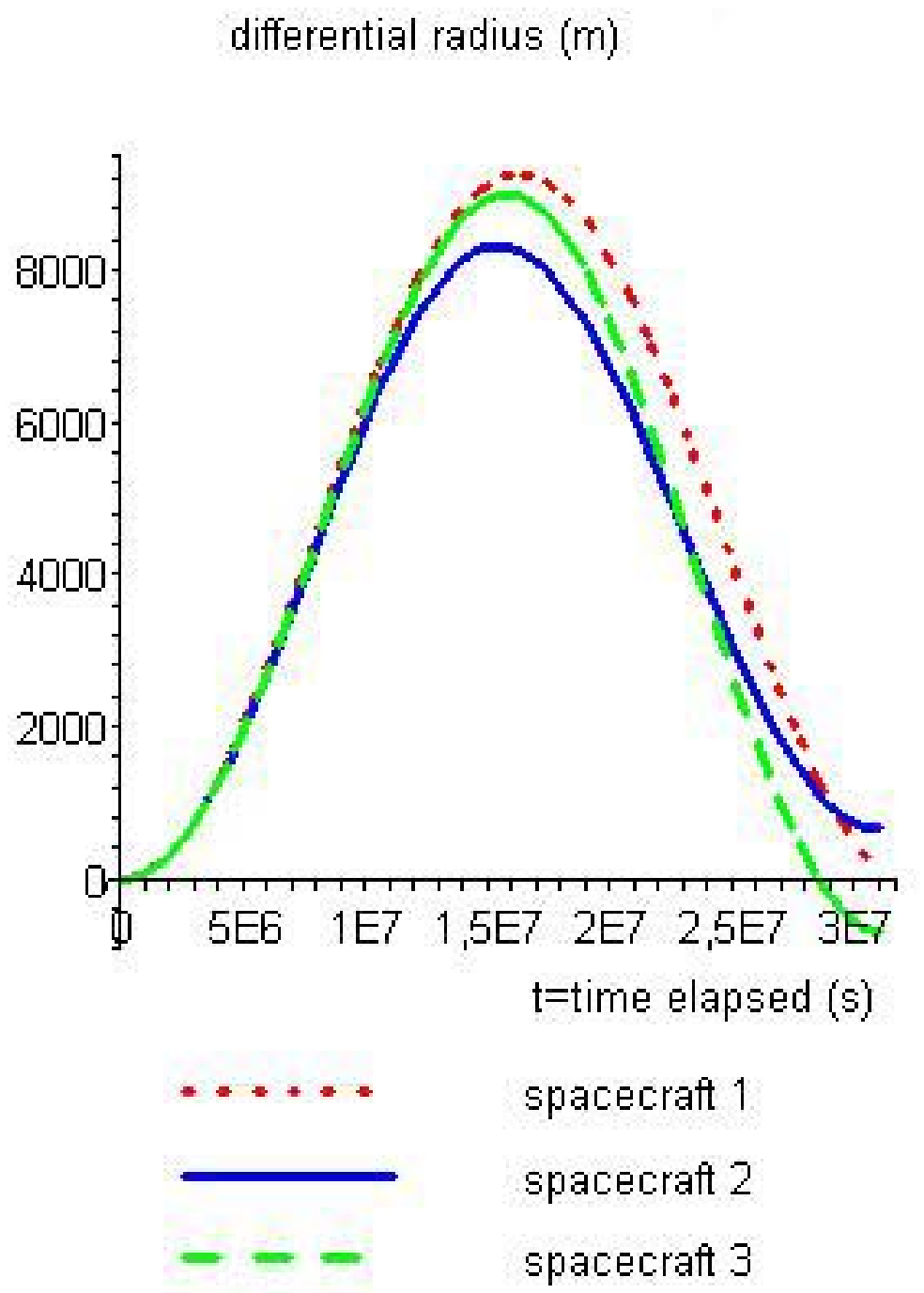}
\end{center}
\caption{Difference between numerical relativistic and classical position
ephemerides for the LISA mission in the \emph{eccentric} ($e_{LISA}\simeq0.0096$)
spherical symmetric case: radial barycentric distance ($\delta r$). }
\label{numeric_diff_relat_kepler_radius_inull}
\end{figure}

\begin{figure}[b]
\begin{center}
\includegraphics[width=0.5\textwidth]{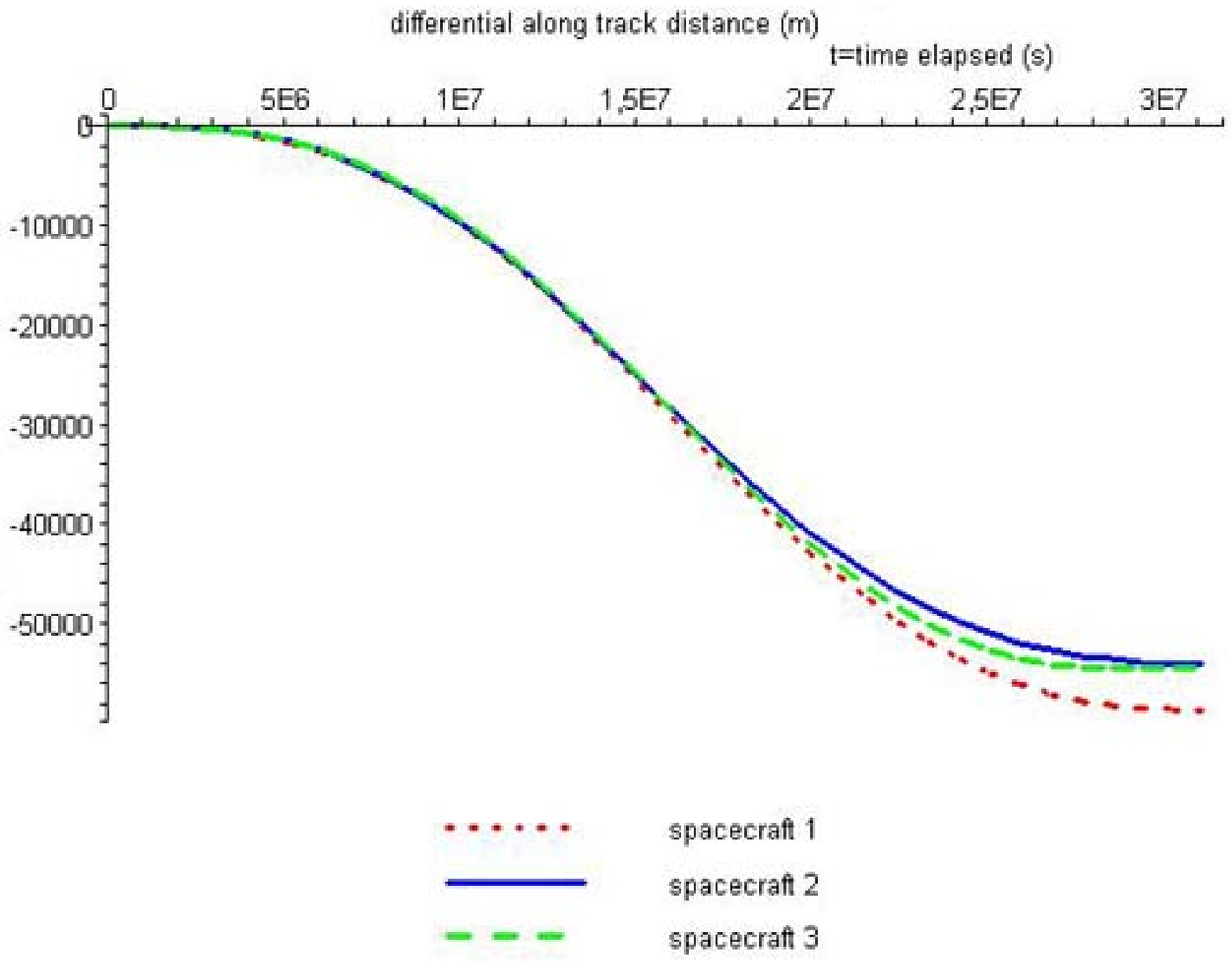}
\end{center}
\caption{Difference between numerical relativistic and classical position
ephemerides for the LISA mission in the \emph{eccentric} ($e_{LISA}\simeq0.0096$)
spherical symmetric case: along track distance ($\delta l$).}
\label{numeric_diff_relat_kepler_along_track_distance_inull}
\end{figure}

From Figure \ref{numeric_diff_relat_kepler_radius_inull}, we see that the maximum difference in radius between numerical relativistic and classical orbits is about $8-9$ km, according to the spacecraft considered. 
From Figure \ref{numeric_diff_relat_kepler_along_track_distance_inull}, 
we see that the along track difference in orbits after one classical
period reaches about $-59$ or $-54$ km, according to the spacecraft considered, for this eccentric spherical symmetric case.\\


\subsubsection{LISA's arm flexing and photon time transfer:}

\begin{figure}[t]
\begin{center}
\includegraphics[width=0.7\textwidth]{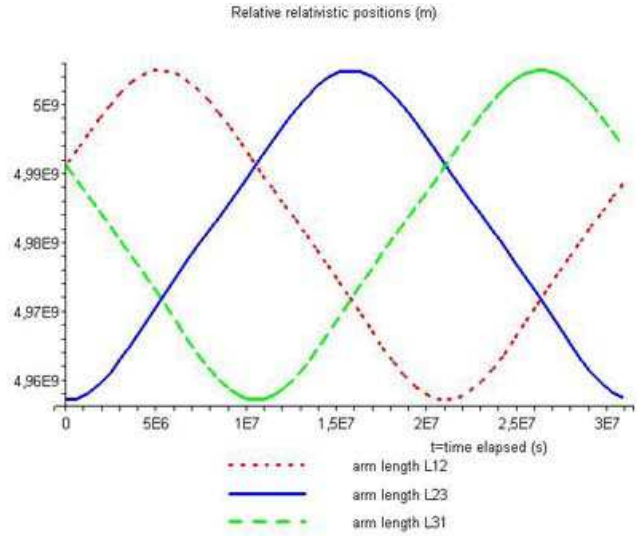}
\end{center}
\caption{Numerical relativistic modelling of LISA breathing in the \emph{eccentric} 
spherical symmetric case ($e_{LISA}\simeq0.0096$): 
relative positions between spacecraft, with $L_{jk}$ the interdistance between spacekraft $j,k=1,2,3$ where $j\neq k$.}
\label{numeric_relativistic_armlength}
\end{figure}

\begin{figure}[t]
\begin{center}
\includegraphics[width=0.5\textwidth]{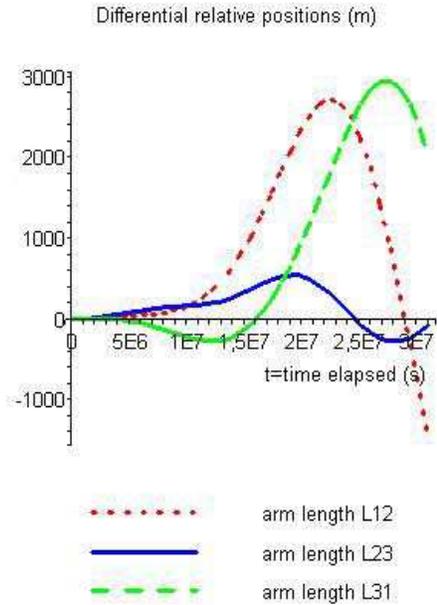}
\end{center}
\caption{Difference between numerical relativistic and classical modelling of LISA breathing 
in the \emph{eccentric} spherical symmetric case ($e_{LISA}\simeq0.0096$): 
difference in relative positions between spacecraft, with $L_{jk}$ the interdistance between spacekraft $j,k=1,2,3$ where $j\neq k$.}
\label{numeric_diff_relat_kepler_armlength}
\end{figure}

Assuming $e=0.0096$ and using the numeric relativistic ephemerides for LISA spacecraft obtained
with the RMI method or that obtained with a classical method, we can
compute the interferometric-arm length $L_{jk}$, that is the interdistance between spacecraft $j$ and $k$.
Over a year, LISA constellation shows some breathing or triangle flexing: the relative position of spacecraft varies as a function of time. 
It is interesting to see that, for the uninclined ($i=0$) eccentric spherical symmetric model, 
the classical approach is wrong by as much as about $4$ km over a one-year mission. However, the true mission has an inclination $i$ such as to minimize the breathing \cite{Na2005}. Figure \ref{numeric_relativistic_armlength} illustrates LISA breathing in the inclined (with the appropriate $i$ given in Section 2) eccentric spherical symmetric case. In that realistic model, the classical approach is wrong by as much as about $3$ km over year of mission, as shown by the residuals (relativistic - classical) relative positions of spacecraft in Figure \ref{numeric_diff_relat_kepler_armlength}. This error translates into a missing $\sim 1\cdot 10^{-5}$ s at zeroth order in $GM/(a\ c^{2})\propto v^{2}/c^{2}$ in photon time transfer ($\stackrel{(0)}{t_{jk}}=L_{jk}/c$) after a year. We recall that, in paper \cite{Chau2005} where the time transfer of photons between LISA spacecraft was studied for a classical LISA orbit, the zeroth order amounted to about $16.7$ s ($5\cdot 10^{6}$ km$/c$, that is the nominal interferometric arm-length, $L$, traveled at the speed of light) with a flexing amplitude of about $0.16$ s ($48000$ km$/c$); the half order amounted to about $3\cdot 10^{- 3}$ s ($960$ km$/c$); and the first order was less than about $1\cdot 10^{-7}$ s ($\leq 30$ m$/c$). Hence, we understand the relevance of relativistic orbit model in the TDI approach, for a coherent modelling of the mission over a few months.

\section{An analytical \emph{development} in eccentricity to check the numerical relativistic 
versus classical orbit model 
}

Let us find an analytical check of the (relativistic - classical) numerical
integration in the eccentric spherical symmetric case, up to first order in $e$
and $GM/c^{2}$.
At the post-Newtonian level the solution is known in terms of osculating elements or other representations (e.g. in \cite{Bru91} or Annex 2 in \cite{So1989}), valid for any eccentricity. 
However, those are implicit solutions (for the radial distance and polar angle) and a further development in eccentricity would be relevant to the LISA mission. 
Indeed, in present LISA literature, orbits and Time Delay Interferometry (TDI) are considered at different levels of approximation, based on a (classical) development in terms of the small eccentricity of the LISA mission 
(an orbit development at a first-order in eccentricity is further assumed by \cite{Arnaud_et_al07}). 
For example, to be ideally a 100 percent efficient in 
removing laser frequency noise and optical bench noises, the TDI combinations from 1st generation TDI algebra assume symmetric and constant (in time) photon propagation time between two LISA spacecraft. This is met only by a rigid motionless constellation model. Hence the need for a 1.5th TDI generation algebra, this time relaxing the symmetry on time-delays. The latter TDI assumptions being met by modeling the constellation as rotating around its center of mass, and around the Sun (without any planet present) in a Keplerian motion at first order in eccentricity. Deviations from this 1st order in eccentricity Keplerian model lead to residual laser frequency and optical bench noise in the TDI combinations, which need to be quantified.
Consequently, the \emph{explicit} general relativistic solution provided in this section as a development at 1PN and first order in eccentricity is useful for the sake of comparison with existing LISA classical models. Our analytical development provides the explicit 
($\delta r_{k}\equiv r_{k\ rel} - r_{k\ cl}$, $\delta \theta_{k}\equiv \theta_{k\ rel} - \theta_{k\ cl}$) relativistic upgrade to the Keplerian 1st order in eccentricity orbit model for LISA such as used by the LMDC \cite{Arnaud_et_al07}. 

To proceed, we first develop the geodesic equation of motion (\ref{non_afine_geodesic_relativistic_BCRS_orbit_equations}) up to the corresponding order in $GM/c^{2}$ in the
BCRS. Writing $\varepsilon ^{l}\equiv x_{rel}^{l}-x_{cl}^{l}$, we find \\

$
\begin{array}{rcl}
\frac{d^{2}x_{rel}^{l}(t)}{d(ct)^{2}} & \simeq  & -\Gamma
_{00}^{l}(x_{rel}^{m}(t)) 
+\Gamma _{00}^{0}(x_{rel}^{m}(t))\cdot \frac{v_{rel}^{l}(t)}{c} \\ 
&  & +2\left[ 
\begin{array}{l}
+\Gamma _{0p}^{0}(x_{rel}^{m}(t))\cdot \frac{v_{rel}^{l}(t)}{c} \\ 
-\Gamma _{0p}^{l}(x_{rel}^{m}(t)) \\ 
-\frac{1}{2}\cdot \Gamma _{qp}^{l}(x_{rel}^{m}(t))\cdot \frac{v_{rel}^{q}(t)}{c}
\end{array}
\right] \cdot \frac{v_{rel}^{p}(t)}{c}
\end{array}
$\medskip \newline

$
\begin{array}{rcl}
\qquad  & \simeq  & -\Gamma _{00}^{l}(x_{cl}^{m}(t))-\varepsilon
^{p}(t)\cdot \frac{\partial \Gamma _{00}^{l}(x_{cl}^{m}(t))}{\partial
x_{cl}^{p}} 
+\Gamma _{00}^{0}(x_{cl}^{m}(t))\cdot \frac{v_{cl}^{l}(t)}{c} \\ 
&  & +2\left[ 
\begin{array}{l}
+\Gamma _{0p}^{0}(x_{cl}^{m}(t))\cdot \frac{v_{cl}^{l}(t)}{c} \\ 
-\Gamma _{0p}^{l}(x_{cl}^{m}(t)) \\ 
-\frac{1}{2}\cdot \Gamma _{qp}^{l}(x_{cl}^{m}(t))\cdot \frac{v_{cl}^{q}(t)}{c}
\end{array}
\right] \cdot \frac{v_{cl}^{p}(t)}{c}
\end{array}
$\medskip \newline
\noindent
where $v^{l}\equiv dx^{l}/dt$ is the velocity of spacecraft at time $t$ in the BCRS. 
Using the analytical developments of Christoffel symbols in the BCRS at the corresponding
order, we can write the difference between the relativistic and classical
orbit accelerations $d^{2}\varepsilon ^{l}/dt^{2}$ as
\begin{eqnarray}
\frac{d^{2}\varepsilon ^{l}}{dt^{2}}+\stackrel{(1)}{A}^{lm}\cdot \varepsilon
^{m} &=&\stackrel{(2)}{A}^{l}  \nonumber \\
&&\text{with}  \nonumber \\
\stackrel{(1)}{A}^{lm} &=&\frac{GM}{r_{cl}^{3}}\cdot \left[ \delta ^{lm}-
\frac{3x_{cl}^{l}x_{cl}^{m}}{r_{cl}^{2}}\right]   \nonumber \\
\stackrel{(2)}{A}^{l} &=&\frac{GM}{r_{cl}^{3}}\cdot \left[ \left( \frac{4GM}
{r_{cl}\ c^{2}}-\frac{v_{cl}^{2}}{c^{2}}\right) x_{cl}^{l}+4\ \frac{v_{cl}^{l}}
{c}\ \frac{v_{cl}^{m}}{c}\ x_{cl}^{m}\right] 
\label{analytical_check_einull_BCRS_orbit_diff_equations}
\end{eqnarray}
\noindent
where $r$ is the coordinate radial distance relative to
the Sun at time $t$ in the BCRS and $(s)$ means that the term considered is of order $s$ in $GM/c^{2}$.\newline
Since we consider a symmetric gravitational field and are interested in the difference between relativistic and classical ephemerides for a given satellite, the inclination $i$ is irrelevant. 
Hence, we choose to work with $i=0$. 
The inclined analytical solution can be obtained by a
simple rotation of the uninclined analytical solution (\ref{analytical_check_einull_BCRS_orbit_tetha_solution}, \ref{analytical_check_einull_BCRS_orbit_radius_solution}). Then of course, 
$z_{cl}=z_{rel}=\varepsilon ^{3}=0$, as well as the corresponding time derivatives.\newline 
Let us further use the set of polar coordinates ($r$, $\theta $) with $x=r\cos \theta $, $y=r\sin \theta $ and  
$z=0 $ to reflect the symmetry of the problem. The above set of equations (\ref
{analytical_check_einull_BCRS_orbit_diff_equations}) becomes 
\begin{equation}
r_{cl}\cdot \delta \stackrel{\bullet \bullet }{\theta }+2\ \stackrel{\bullet 
}{r}_{cl}\cdot \delta \stackrel{\bullet }{\theta }+2\ \stackrel{\bullet }
{\theta }_{cl}\cdot \delta \stackrel{\bullet }{r}+\stackrel{\bullet \bullet }
{\theta }_{cl}\cdot \delta r=\frac{4GM}{c^{2}}
\frac{\stackrel{\bullet }{r}_{cl}}{r_{cl}}\stackrel{\bullet }{\theta }_{cl}
\label{analytical_check_einull_BCRS_orbit_diff_equations_theta}
\end{equation}
\begin{eqnarray}
\delta \stackrel{\bullet \bullet }{r}-\left[ \stackrel{\bullet }{\theta }_{cl}^{2}
+\frac{2GM}{r_{cl}^{3}}\right] \cdot \delta r-2\ r_{cl}\stackrel{\bullet }
{\theta }_{cl}\cdot \delta \stackrel{\bullet }{\theta } &=&\frac{4GM}
{c^{2}}\cdot \left[ \frac{GM}{r_{cl}^{3}}-\frac{v_{cl}^{2}}{4\ r_{cl}^{2}}
+\frac{\stackrel{\bullet }{r}_{cl}^{2}}{r_{cl}^{2}}\right]   \nonumber \\
&&  \label{analytical_check_einull_BCRS_orbit_diff_equations_r}
\end{eqnarray}
\noindent
where $\delta \theta \equiv \theta _{rel}-\theta _{cl}$, $\delta r\equiv
r_{rel}-r_{cl}$ and $\stackrel{\bullet }{*}\equiv d*/dt$.\newline
Using Kepler's orbital motion equations ($r_{cl}=a(1-e^{2})/(1+e\cos \theta
_{cl})$, $r_{cl}^{2}\stackrel{\bullet }{\theta }=\sqrt{GM\ a\ (1-e^{2})}$, 
$v_{cl}=\sqrt{GM\ (2/r_{cl}-1/a)}$), we can check that equations (\ref
{analytical_check_einull_BCRS_orbit_diff_equations_theta}) and 
(\ref{analytical_check_einull_BCRS_orbit_diff_equations_r}) lead to two first integrals of the motion: 
\begin{equation}
r_{cl}^{2}\cdot \delta \stackrel{\bullet }{\theta }+2\ r_{cl}\cdot 
\stackrel{\bullet }{\theta }_{cl}\cdot \delta r+\frac{4GM}{r_{cl}\ c^{2}}\sqrt{GM\ a\
(1-e^{2})}=C_{k}  \label{analytical_check_einull_BCRS_orbit_first_integral_J}
\end{equation}
\begin{eqnarray}
\stackrel{\bullet }{r}_{cl}\cdot \delta \stackrel{\bullet }{r}+\left[
r_{cl}\cdot \stackrel{\bullet }{\theta }_{cl}^{2}+\frac{GM}{r_{cl}^{2}}\right] 
\cdot \delta r+r_{cl}^{2}\stackrel{\bullet }{\theta }_{cl}\cdot
\delta \stackrel{\bullet }{\theta }+\frac{GM}{c^{2}}\left[ -\frac{GM}
{r_{cl}^{2}}+3\frac{v_{cl}^{2}}{r_{cl}}\right]  &=&D_{k}  \nonumber \\
&&  \label{analytical_check_einull_BCRS_orbit_first_integral_K}
\end{eqnarray}
\noindent
Those can be traced back to the relativistic angular
momentum and energy integral 
resulting from the spherical symmetry.
\newline
Owing to our choice of identical positions and velocities of
spacecraft at initial time for both the classical and the relativistic
orbit models, $(\delta \theta ,\delta r,\delta \stackrel{\bullet }
{\theta },\delta \stackrel{\bullet }{r})_{t_{0}}=(0,0,0,0)$. Hence the integration
constants are 
\begin{eqnarray*}
C_{k} &=&4\frac{GM}{c^{2}}\frac{\sqrt{GM\ a\ (1-e^{2})}}{r_{k\ cl\ 0}} \\
D_{k} &=&\frac{GM}{c^{2}}\ \left[ 3\frac{v_{k\ cl\ 0}^{2}}{r_{k\ cl\ 0}}
-\frac{GM}{r_{k\ cl\ 0}^{2}}\right] 
\end{eqnarray*}
\noindent
with $r_{k\ cl\ 0}\equiv r_{k\ cl}(t_{0})$ and $v_{k\ cl\ 0}\equiv v_{k\ cl}(t_{0})$ given by Kepler's orbital
equation of motion at initial time with respect to the initial conditions of
a given spacecraft $k=1,2$ or $3$. Equations (\ref
{analytical_check_einull_BCRS_orbit_first_integral_J}, \ref
{analytical_check_einull_BCRS_orbit_first_integral_K}) provide a first check
of the numerical results of Sections \ref{numerical_results_einull} and \ref{numerical_results_inull} in the spherical symmetric approximation. We note that for a circular orbit
of reference ($n_{cl}=\stackrel{\bullet }{\theta }_{cl}$, $r_{cl}=a$, 
$\stackrel{\bullet }{r}_{cl}=0$), $C_{k}$ and the third term of the
left-hand-side of (\ref{analytical_check_einull_BCRS_orbit_first_integral_J}) 
cancel; while $D_{k}$ and the fourth term of the left-hand-side of (\ref
{analytical_check_einull_BCRS_orbit_first_integral_K}) cancel... leading to
the same identical first integral: $\delta \stackrel{\bullet }{l}=-2\ n_{cl}\
\delta r$, where $\delta l\equiv  r_{rel}\cdot \delta \theta$.

We now develop the differential system ((\ref
{analytical_check_einull_BCRS_orbit_diff_equations_theta}) and (\ref
{analytical_check_einull_BCRS_orbit_diff_equations_r}); or, which is easier, 
(\ref{analytical_check_einull_BCRS_orbit_first_integral_J}) and (\ref
{analytical_check_einull_BCRS_orbit_first_integral_K})), up to first order in 
$e$ using Kepler's equations of motion at first order in $e$: 
\begin{eqnarray*}
\frac{C_{k}}{a\ n_{cl}} &=&\left( 1-2\ e\cos \theta _{cl}\right) \cdot
\delta \stackrel{^{\prime }}{l}+2\ \left( 1+e\cos \theta _{cl}\right)
\cdot \delta r+\frac{4GM}{c^{2}}\left( 1+e\cos \theta _{cl}\right)  \\
\frac{D_{k}}{a\ n_{cl}^{2}} &=&\delta \stackrel{^{\prime }}{l}+e\sin
\theta _{cl}\cdot \delta \stackrel{^{\prime }}{r}+\left( 2+5\ e\cos
\theta _{cl}\right) \cdot \delta r+\frac{GM}{c^{2}}\left( 2+7\ e\cos
\theta _{cl}\right) 
\end{eqnarray*}
\noindent
with $\stackrel{\prime }{*}\equiv d*/d(n_{cl}t)$. 
To find solutions to the above differential system, we use the theory of
perturbation around null eccentricity. We find 
\begin{eqnarray}
\delta \theta _{k} &\simeq &\stackrel{[0]}{\delta \theta }_{k}+\stackrel{[1]}
{\delta \theta }_{k}  \nonumber \\
&&\text{with}  \nonumber \\
\stackrel{\lbrack 0]}{\delta \theta }_{k} &=&-6\frac{GM}{a\ c^{2}}\left\{ 
\begin{array}{l}
+n_{cl}t-\cos \left( n_{cl}t_{kp}\right) \sin \left( n_{cl}(t-t_{kp})\right) 
\\ 
-\sin \left( n_{cl}t_{kp}\right) \cos \left( n_{cl}(t-t_{kp})\right) 
\end{array}
\right\}   \nonumber \\
\stackrel{\lbrack 1]}{\delta \theta }_{k} &=&+e\frac{GM}{a\ c^{2}}\left\{ 
\begin{array}{l}
+2\sin \left( n_{cl}t_{kp}\right) -21\cos \left( n_{cl}t_{kp}\right) n_{cl}t
\\ 
-18\ n_{cl}t\cos \left( n_{cl}(t-t_{kp})\right)  \\ 
+22\cos \left( n_{cl}t_{kp}\right) \sin \left( n_{cl}t_{kp}\right) \cos
\left( n_{cl}(t-t_{kp})\right)  \\ 
+\left\{ 2+22\cos ^{2}\left( n_{cl}t_{kp}\right) \right\} \sin \left(
n_{cl}(t-t_{kp})\right)  \\ 
+15\sin \left( n_{cl}t_{kp}\right) \cos ^{2}\left( n_{cl}(t-t_{kp})\right) 
\\ 
+15\cos \left( n_{cl}t_{kp}\right) \sin \left( n_{cl}(t-t_{kp})\right) \cos
\left( n_{cl}(t-t_{kp})\right) 
\end{array}
\right\}   \nonumber \\
&&  \label{analytical_check_einull_BCRS_orbit_tetha_solution}
\end{eqnarray}
\begin{eqnarray}
\delta r_{k} &\simeq &\stackrel{[0]}{\delta r}_{k}+\stackrel{[1]}{\delta r}_{k}  \nonumber \\
&&\text{with}  \nonumber \\
\stackrel{\lbrack 0]}{\delta r}_{k} &=&+3\frac{GM}{c^{2}}\left\{ 
\begin{array}{l}
+1-\cos \left( n_{cl}t_{kp}\right) \cos \left( n_{cl}(t-t_{kp})\right)  \\ 
+\sin \left( n_{cl}t_{kp}\right) \sin \left( n_{cl}(t-t_{kp})\right) 
\end{array}
\right\}   \nonumber \\
\stackrel{\lbrack 1]}{\delta r}_{k} &=&+e\frac{GM}{c^{2}}\left\{ 
\begin{array}{l}
+20\cos \left( n_{cl}t_{kp}\right) -9\ n_{cl}t\sin \left(
n_{cl}(t-t_{kp})\right)  \\ 
-\left\{ 3+11\cos ^{2}\left( n_{cl}t_{kp}\right) \right\} \cos \left(
n_{cl}(t-t_{kp})\right)  \\ 
+11\cos \left( n_{cl}t_{kp}\right) \sin \left( n_{cl}t_{kp}\right) \sin
\left( n_{cl}(t-t_{kp})\right)  \\ 
-6\cos \left( n_{cl}t_{kp}\right) \cos ^{2}\left( n_{cl}(t-t_{kp})\right) 
\\ 
+6\sin \left( n_{cl}t_{kp}\right) \sin \left( n_{cl}(t-t_{kp})\right) \cos
\left( n_{cl}(t-t_{kp})\right) 
\end{array}
\right\} \nonumber   \\
&&  \label{analytical_check_einull_BCRS_orbit_radius_solution}
\end{eqnarray}
\noindent
where $[s]$ means that the term considered is of order $s$ in $e$. At zeroth order in $e$, 
those results correspond to the circular classical orbit of reference case.\newline
Expressions (\ref{analytical_check_einull_BCRS_orbit_tetha_solution}) and (\ref{analytical_check_einull_BCRS_orbit_radius_solution}) 
can be easily transposed in terms of (relativistic -
classical) positions ($\delta x_{k}$, $\delta y_{k}$) and related
(relativistic - classical) coordinate velocities ($\delta \stackrel{\bullet 
}{x}_{k}$, $\delta \stackrel{\bullet }{y}_{k}$) using 
\[
\left\{ 
\begin{array}{l}
x_{k}=r_{k}\cos \theta _{k} \\ 
y_{k}=r_{k}\sin \theta _{k}
\end{array}
\right. \Rightarrow \left\{ 
\begin{array}{l}
\delta x_{k}=\cos \theta _{k}\cdot \delta r_{k}-r_{k}\sin \theta _{k}\cdot
\delta \theta _{k} \\ 
\delta y_{k}=\sin \theta _{k}\cdot \delta r_{k}+r_{k}\cos \theta _{k}\cdot
\delta \theta _{k}
\end{array}
\right. . 
\]


\subsection{Circular classical reference orbit case:}

\begin{figure}[t]
\begin{center}
\includegraphics[width=0.5\textwidth]{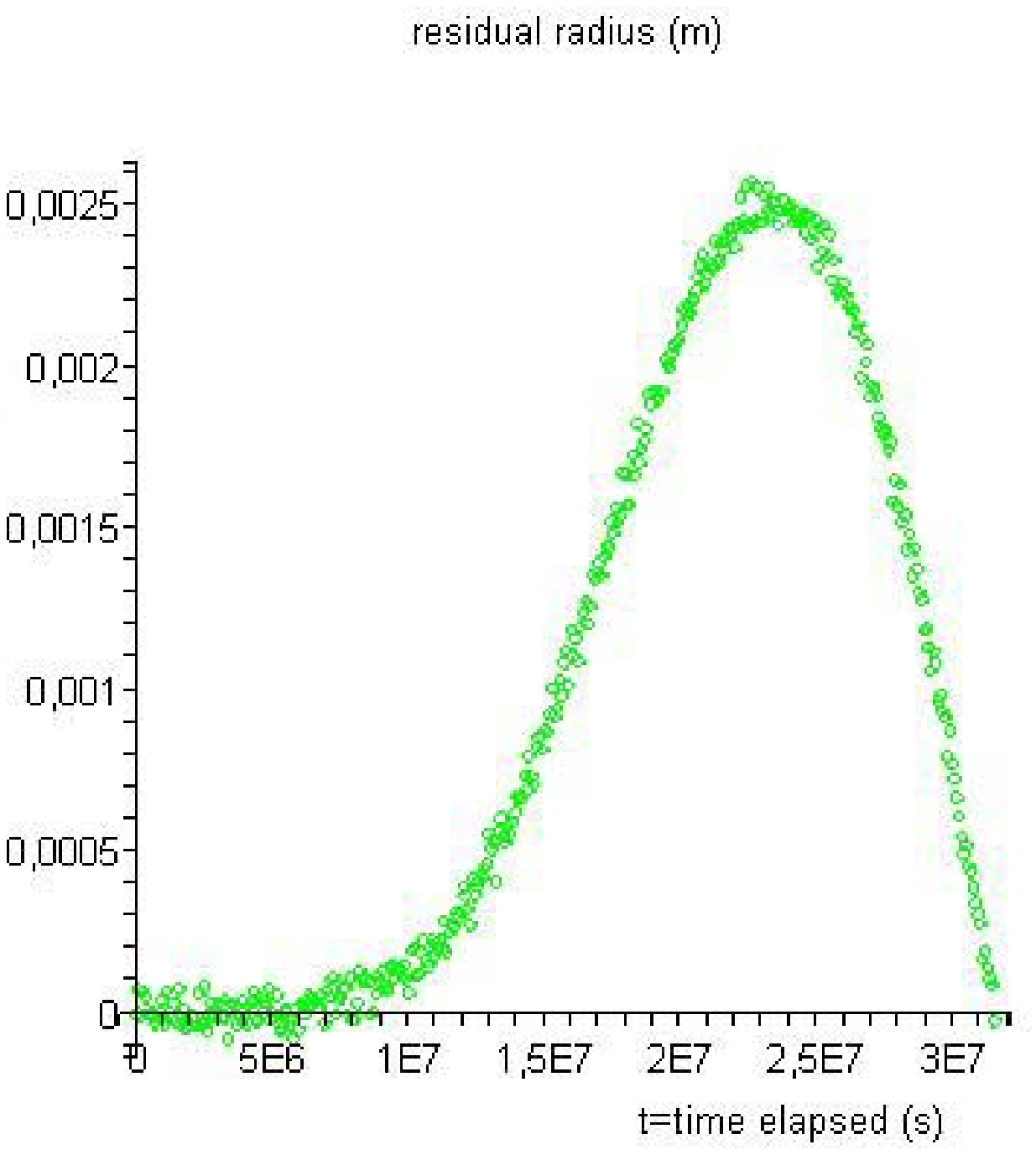}
\end{center}
\caption{Residuals between the numerical (relativistic -
classical) position ephemerides and the corresponding analytical development 
for the LISA mission in the \emph{circular} ($e=0$) spherical
symmetric case: radial distance ($\delta r $).}
\label{numeric_analytic0_residual_diff_relat_kepler_radius_einull}
\end{figure}

\begin{figure}[t]
\begin{center}
\includegraphics[width=0.5\textwidth]{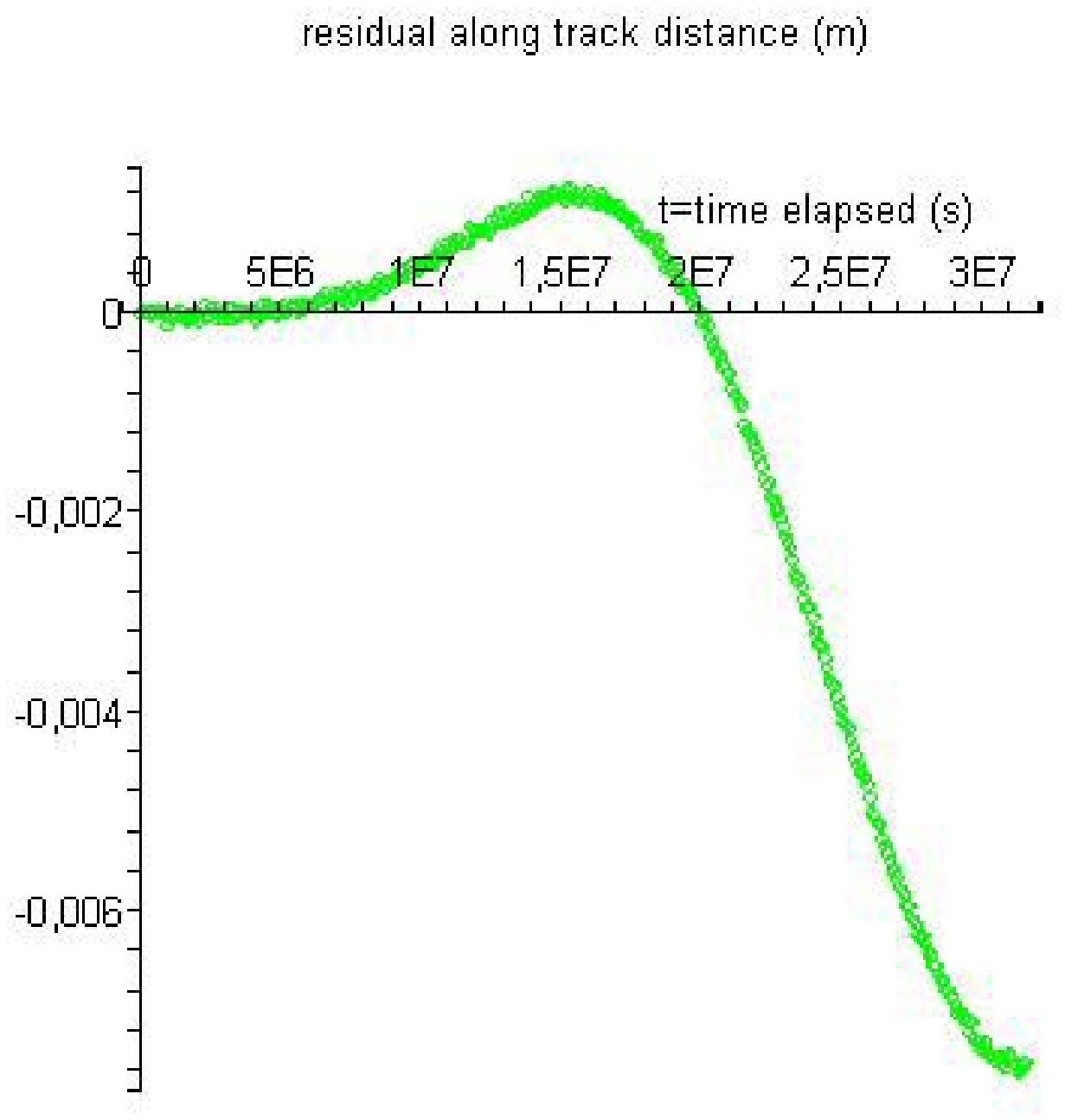}
\end{center}
\caption{Residuals between the numerical (relativistic -
classical) position ephemerides and the corresponding analytical development for the LISA mission in the \emph{circular} ($e=0$) spherical symmetric case: along track distance ($\delta l $).}
\label{numeric_analytic0_residual_diff_relat_kepler_along_track_distance_einull}
\end{figure}

Expressions (\ref{analytical_check_einull_BCRS_orbit_tetha_solution}) 
and (\ref{analytical_check_einull_BCRS_orbit_radius_solution}) with $e=0$ match perfectly the
numerical results for the circular spherical symmetric case presented in Section \ref{numerical_results_einull}, 
up to first order in $GM/c^{2}$. Residuals between RMI approach and this analytical check for the circular spherical symmetric case reach about 1 cm in x-y-positions or along track distance and a few millimeters in radius (Figures \ref{numeric_analytic0_residual_diff_relat_kepler_radius_einull} and \ref{numeric_analytic0_residual_diff_relat_kepler_along_track_distance_einull}).\\
A dimensional analysis leads to an order of magnitude for the difference between classical and
relativistic barycentric positions of spacecraft of about $GM/(a c^2)\cdot 2\pi a\simeq $10 km for a one year simulation. 
Our numerical native relativistic approach shows that classical modelling can be wrong 
by as much as about 50 km, in terms of barycentric coordinates (x,y,z) and along track distance, 
over one year. It is interesting to point out that this is nearly one order of magnitude larger than estimated with a dimensional analysis. 
The numerical results are confirmed by the more cautious analytical developpements presented above.


\subsection{Eccentric classical reference orbit case}

\begin{figure}[t]
\begin{center}
\includegraphics[width=0.5\textwidth]{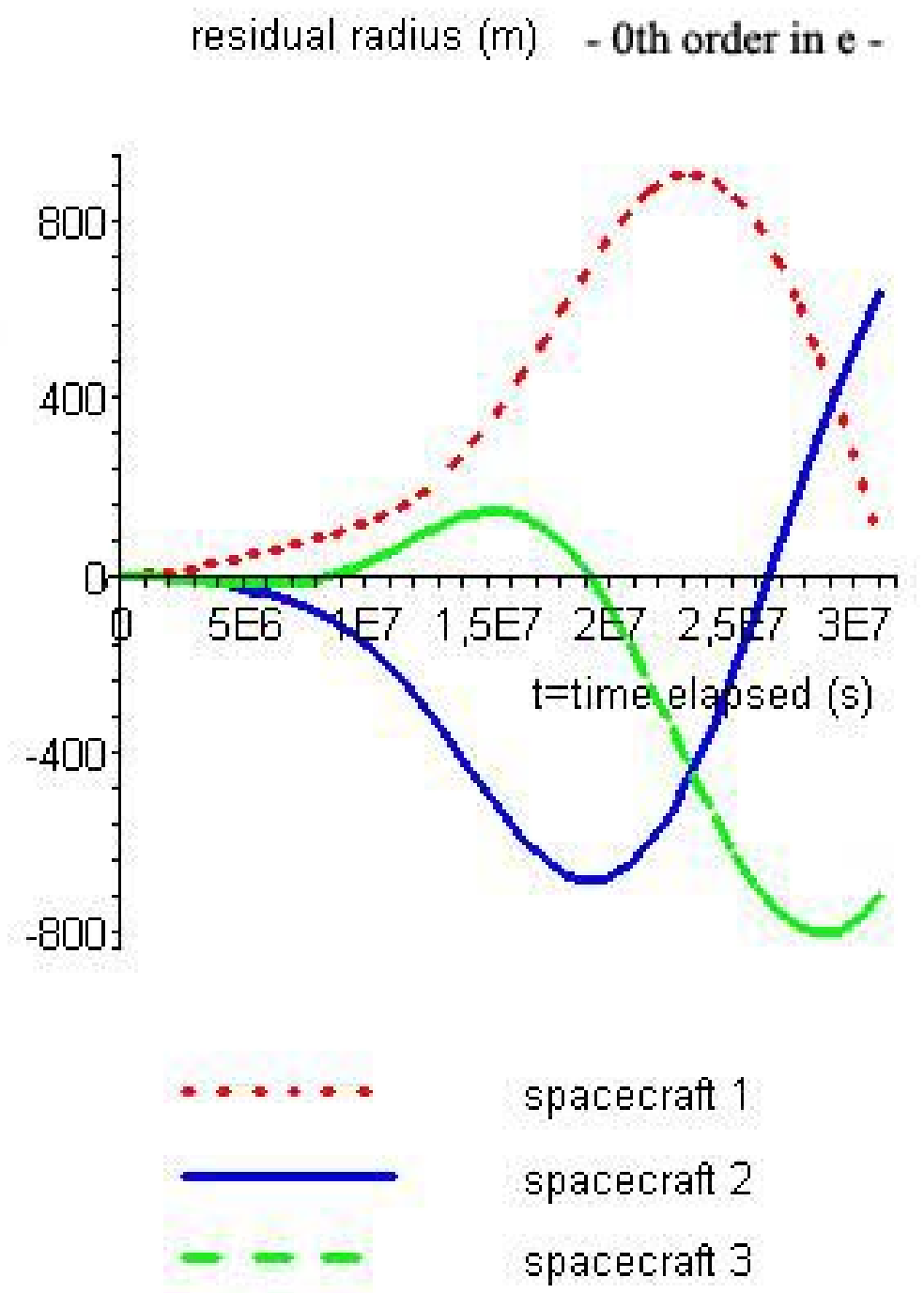}
\end{center}
\caption{Residuals between the numerical (relativistic -
classical) position ephemerides and the corresponding analytical development at 0th order in $e$
for the LISA mission in the \emph{eccentric} ($e_{LISA}\simeq0.0096$) spherical
symmetric case: radial distance ($\delta r $).}
\label{numeric_analytic0_residual_diff_relat_kepler_radius_inull}
\end{figure}

\begin{figure}[t]
\begin{center}
\includegraphics[width=0.5\textwidth]{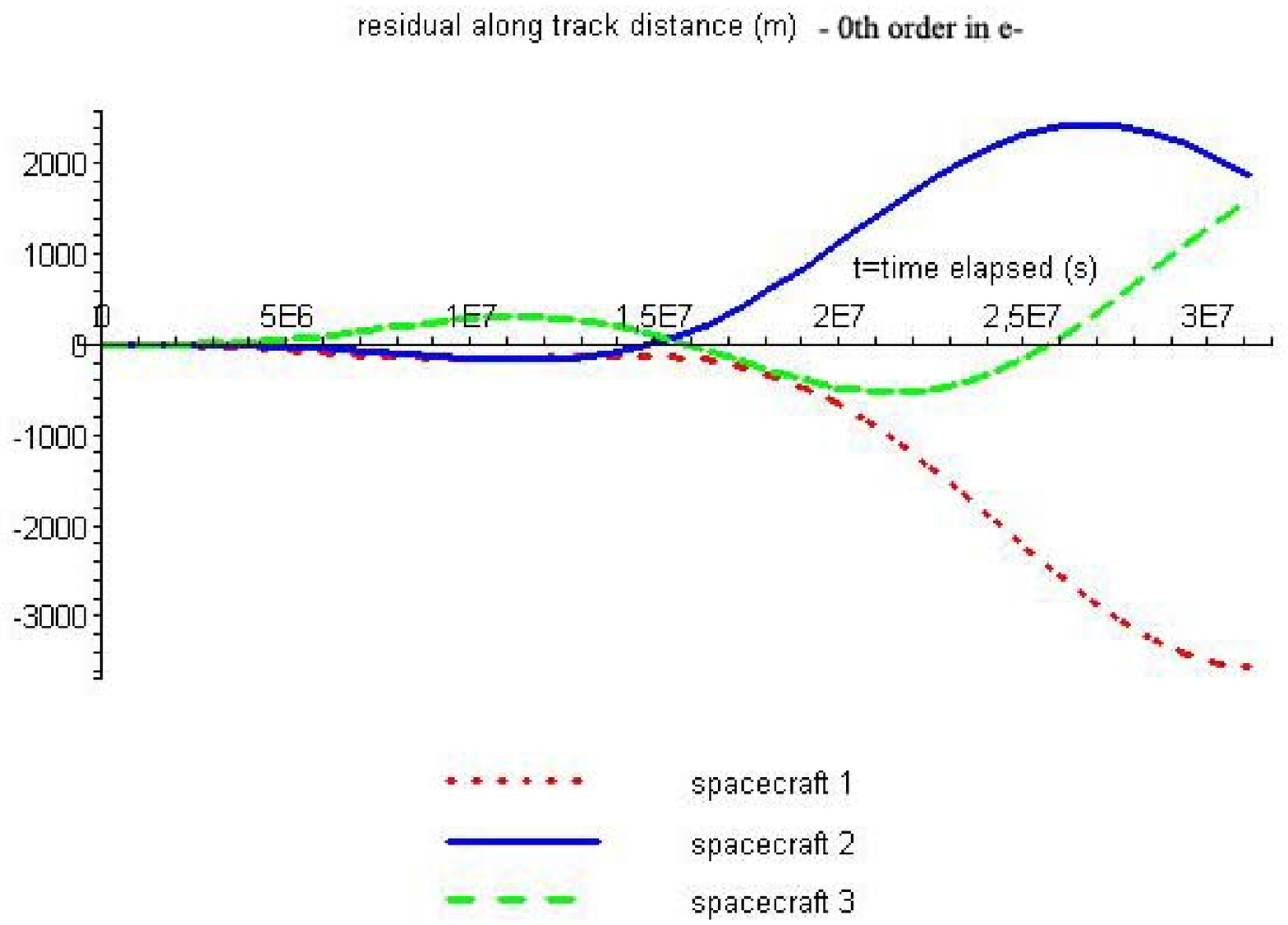}
\end{center}
\caption{Residuals between the numerical (relativistic -
classical) position ephemerides and the corresponding analytical development at 0th order
in $e$ for the LISA mission in the \emph{eccentric} ($e_{LISA}\simeq0.0096$)
spherical symmetric case: along track distance ($\delta l $).}
\label{numeric_analytic0_residual_diff_relat_kepler_along_track_distance_inull}
\end{figure}

\begin{figure}[t]
\begin{center}
\includegraphics[width=0.5\textwidth]{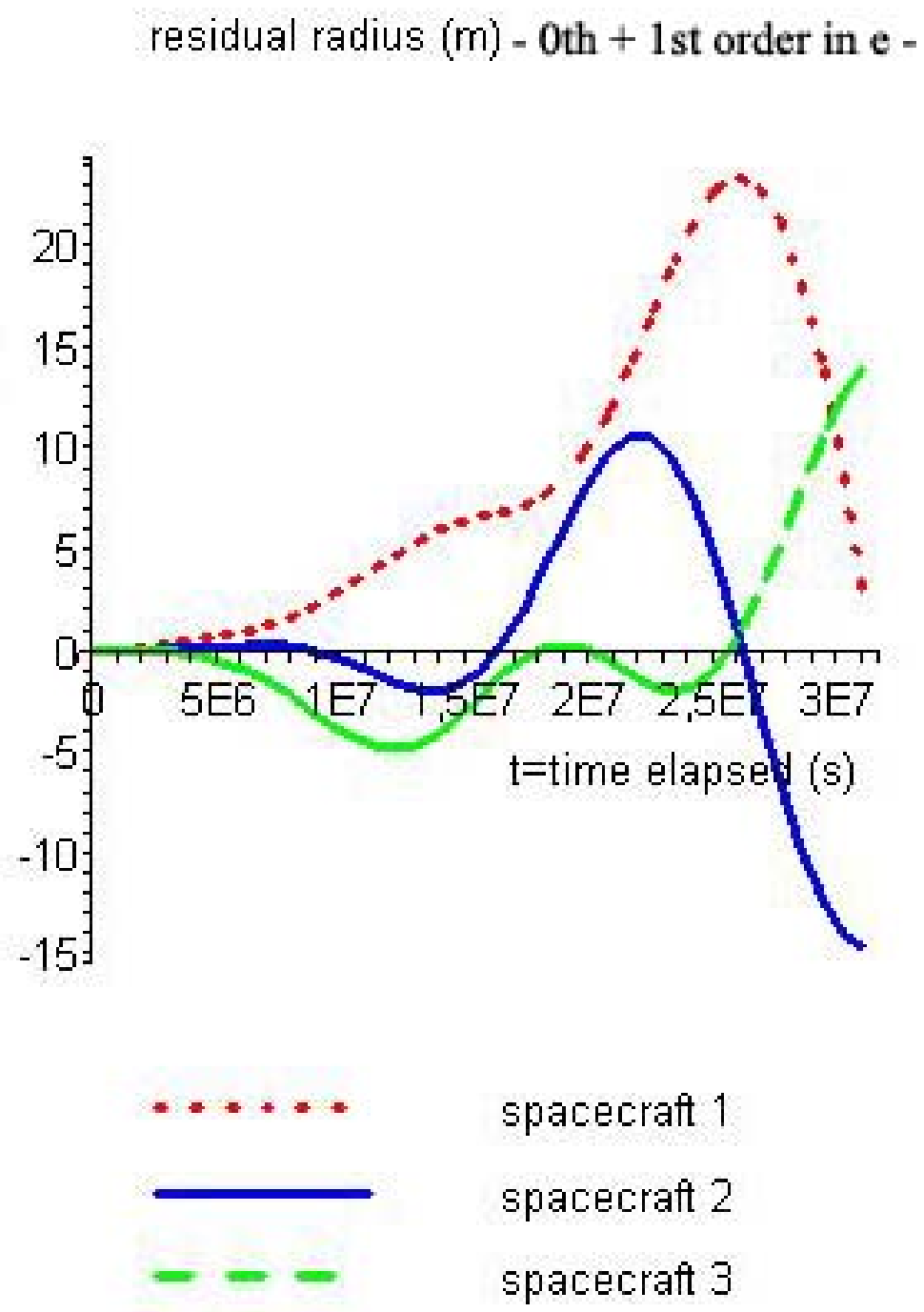}
\end{center}
\caption{Residuals between the numerical (relativistic -
classical) position ephemerides and the corresponding analytical development up to 1st
order in $e$ for the LISA mission in the \emph{eccentric} ($e_{LISA}\simeq0.0096$)
spherical symmetric case: radial distance ($\delta r $).}
\label{numeric_analytic01_residual_diff_relat_kepler_radius_inull}
\end{figure}

\begin{figure}[t]
\begin{center}
\includegraphics[width=0.5\textwidth]{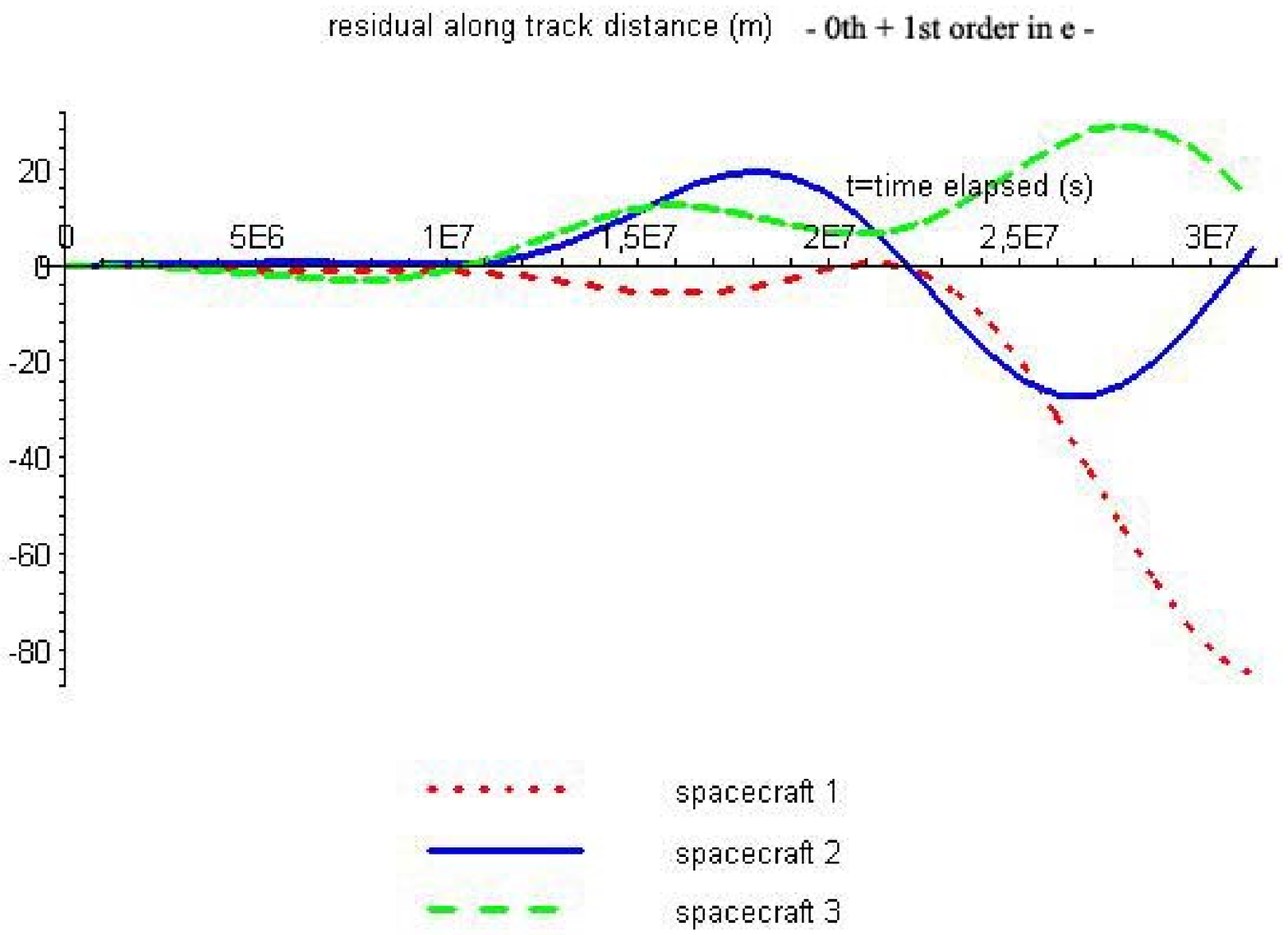}
\end{center}
\caption{Residuals between the numerical (relativistic -
classical) position ephemerides and the corresponding analytical development up to 1st
order in $e$ for the LISA mission in the \emph{eccentric} ($e_{LISA}\simeq0.0096$)
spherical symmetric case: along track distance ($\delta l $).}
\label{numeric_analytic01_residual_diff_relat_kepler_along_track_distance_inull}
\end{figure}

Expressions (\ref{analytical_check_einull_BCRS_orbit_tetha_solution}) 
and (\ref{analytical_check_einull_BCRS_orbit_radius_solution}) with
orbital elements corresponding to LISA's ($e=0.0096$) but $i=0$ match the
numerical results for the eccentric spherical symmetric case presented in
Section \ref{numerical_results_inull}, up to first order in $e$ and in $GM/c^{2}$.\newline
Residuals between the RMI approach and this analytical check \emph{at zeroth order in $e$}, for the eccentric
spherical symmetric case, reach up to about $+840$, $\pm 540$ or $-800$ m in radial 
distance and about $-3600$, $+2400$ or $+1600$ m in along track distance, 
for spacecraft $k=1,2$ or $3$ respectively, over a year 
(Figures \ref{numeric_analytic0_residual_diff_relat_kepler_radius_inull} and \ref{numeric_analytic0_residual_diff_relat_kepler_along_track_distance_inull}).\newline
When the analytical check for the eccentric spherical symmetric case is considered 
\emph{up to first order in $e$}, the residuals reach up to about $+24$, $-15$ or $+14$ m in radius 
and about $-85$, $-25$ or $+32$ m in along track distance, for spacecraft $k=1,2$ or $3$ respectively, over a year (Figures \ref{numeric_analytic01_residual_diff_relat_kepler_radius_inull} and \ref{numeric_analytic01_residual_diff_relat_kepler_along_track_distance_inull}). 
Residuals between the RMI numerical analysis (implicitly containing all orders in $e$) and the analytical development (up to first order in $e$, equations (\ref{analytical_check_einull_BCRS_orbit_tetha_solution}) and (\ref{analytical_check_einull_BCRS_orbit_radius_solution}) ) are bound to be larger for 
space missions with larger eccentricities than that of LISA's ($e_{LISA}\simeq 0.0096$). This shows the limits of the analytical development even for an \emph{eccentric} model with a simple spherical symmetric gravitational field.
And going to higher orders in $e$ increases the number of terms in expressions (\ref{analytical_check_einull_BCRS_orbit_tetha_solution}) and (\ref{analytical_check_einull_BCRS_orbit_radius_solution}) drastically, as illustrated by the 0-th and 1-st order contributions. 
Non symmetric cases such as in presence of planets, with a central body which is non spherical or has a spin, 
are even much more complex to handle analytically. 
On the opposite, the RMI approach, which is exact in terms of $e$, implicit in terms of spin, flattening or planets, via the metric, is very flexible. Indeed, RMI also runs when the spherical symmetry is broken, since 
solar spin, multipolar development of the solar mass, point-like planets can be introduced in the metric and hence be coherently taken into account in numerical ephemerides produced for LISA via that approach.

\section{Conclusions}

\qquad The aim of the present paper was to illustrate how the 
Relativistic Motion Integrator (RMI) can be used to provide a relativistic numerical satellite (or test-body) 
propagator for space missions; and to quantify relativistic effects when a comparison is 
made with a classical corresponding model.
 
As an illustration of RMI and to validate the method, we chose the space interferometer LISA, modelled in the Barycentric
Coordinate Reference System (BCRS) in the gravitational field of a spherical
non-rotating Sun, without planets (the spherical symmetric case). 
We compared the numerical relativistic ephemeris (propagated daily positions and velocities of
each spacecraft) obtained with RMI to the ones obtained with a classical numerical model
with identical initial conditions in terms of positions and velocities.  
The
(relativistic - classical) BCRS position obtained seemed a priori large, up
to a few tenth kilometers, i.e. more or less 5 or 6 times the estimate obtained from a rapid dimensional 
analysis.\\
However, we made a careful analytical analysis: 
analytical expressions (up to first order in $GM/c^{2}$, 
with $G$, Newton's constant, $M$, the Sun's mass and $c$, the speed of light in vacuum) 
of two first integrals of the problem and an analytical development of
(relativistic - classical) BCRS along track and radial distances up to first
order in eccentricity $e$ and in $GM/c^{2}$. The analytical
developments with orbital elements corresponding to LISA's confirmed the
numerical results obtained and validates the RMI approach. The difference
between the RMI numerical approach, based on the exact relativistic equation
of motion with respect to the BCRS metric 
(which is up to second order in $GM/c^{2}$ in the IAU2000 resolutions) 
for a spherical non-rotating Sun, 
and the analytical development are of order $e^{2} \cdot GM/c^{2}$.\newline
Hence, for LISA, we have shown that, when the classical orbit of reference is eccentric
with $e_{LISA}\simeq0.0096$, the difference between relativistically and
classically modelled radial distance reaches up to a maximum of about $8-9$
km during a one-year mission. After one year (i.e. one classical period), 
the difference in orbits in terms of radial distance can be as much as about $680$ m and 
along track difference is about $54-59$ km according to the spacecraft considered. \\
Errors in LISA satellite orbit may have consequences when modelling LISA's arm flexing 
for the sake of interferometry. We showed that a relativistic orbit model is relevant when studying photon time transfer needed in the TDI method. 
Using a classical orbit model contributes to an error of about $10^{-5}$ s ($\approx 3$ km$/c$) in photon time transfer over a year. 
The TDI method is the crucial pre-processing of LISA data, before even trying a given strategy to detect any gravitational wave signal. \\
Since the orders of
magnitude of $(a_{k},e_{k},i_{k})_{k=1,2,3}$ used in Hughes 
orbit model for LISA's three spacecraft \cite{Hu2005} are the same as the
ones chosen here, the same conclusions will apply in Hughes'case. \newline
Note that, in the present paper, we did not aim at a complete model of the LISA detector, but some of the above result might be interesting when building 
a LISA simulator.

Our present study also shows that while the analytical development soon reaches its limits, 
the strength of the RMI approach is that it also
runs when the spherical symmetry is broken ($i\neq 0$, non-spherical Sun,
rotating Sun, with planets), cases much more complex to model analytically.
Indeed, a solar spin or multipolar development of the solar mass (solar 
$J_{2}$) or point-like planets can be introduced in the metric and hence be
coherently taken into account in numerical ephemerides produced for LISA\
via the RMI approach.
The point is to use a metric with a sufficiently high order of development in $1/c^{2}$, 
so as to include all the classical and relativistic effects relevant to the precision of 
the space mission considered. The IAU 2000 BCRS metric models coherently, for LISA 
and other space missions, the 
action of the Sun and planets at a relativistic level. 

Finally, the RMI approach can be applied to other space
missions, whether barycentric or planetocentric.\pagebreak

\clearpage


\section{FIGURES}

\label{figures}

%
%
%
%
%

\pagebreak

\clearpage

\appendix
\section{Numerical estimate of the Christoffel Symbols}
\label{numerical_integration_precision}

\begin{figure}
 \centering
 \includegraphics[width=.8\textwidth]{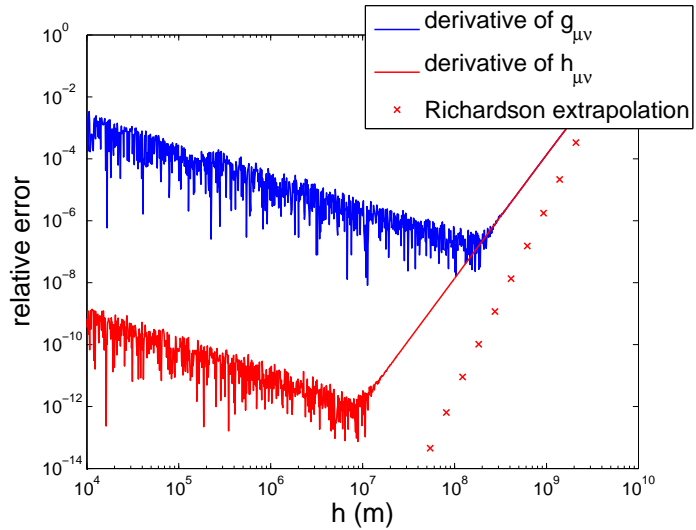}
 \caption{Representation of the relative precision of $g_{tt,x}$ for one point of the LISA orbit. The relative precision of $g_{tt,x}$ and $h_{tt,x}$ are represented as function of the discretisation step, $h$. The Richardson extrapolation is also represented for a factor $k=1.5$ (12 iterations are represented). }
 \label{figDer}
\end{figure}

Within a numerical integration of the relativistic equations of motion, one has to carefully check the numerical accuracy. In this section, we show that the numerical errors are smaller than the order of magnitude of the relativistic effects. \\
In order to integrate equation (\ref{relativistic_BCRS_orbit_equations}), we need to evaluate numerically the Christoffel Symbols
\begin{equation}
 \Gamma_{\mu\nu}^{\alpha}=\frac{1}{2}g^{\alpha\beta}\left(g_{\beta \nu,\mu}+g_{\mu\beta,\nu}-g_{\mu\nu,\beta}\right)
\end{equation}
where $f_{,x}=\frac{\partial f}{\partial x}$ and the matrix $g^{\alpha\beta}$ is the inverse of the matrix  $g_{\alpha\beta}$ owing to expression (\ref{cov_contra_metric_components}). 
We need to evaluate numerically the derivative, $g_{\mu\nu,\beta}$, of the metric components. 
The derivative is computed using an estimation of order 4~\cite{KinChe2002}
\begin{equation}
 D_h(x)= \frac{f(x-2h)-8f(x-h)+8f(x+h)-f(x+2h)}{12h}
\end{equation}
\begin{equation}
 \text{with} \quad f'(x)=D_h(x)+\mathcal O(h^4)
\end{equation}
As can be seen in Figure \ref{figDer}, one needs to choose the discretisation step size, $h$, very carefully. 
For large $h$, the discretisation error is important ($\propto h^4$) but for small $h$, the roundoff error increases ($\propto 1/h$).

In order to increase the precision of the derivative, it is usefull to derive $h_{\mu\nu}=g_{\mu\nu}-\eta_{\mu\nu}$, where $\eta_{\mu\nu}$ is the Minkowsky metric, instead of $g_{\mu\nu}$ as it is more stable from a numerical point of view (Figure \ref{figDer}). 

It is also interesting to use Richardson extrapolation \cite{RiGau1927}. This requires two estimations of order 4 ($D_{0,0}=D_h(x)$  and $D_{1,0}=D_{h/k}(x)$ where $k$ is a real factor) to construct a new estimation of order 8:
\begin{equation}
 D_{1,1}= \frac{k^4D_{1,0}-D_{0,0}}{k^4-1}
\end{equation}
In practice, the factor $k$ is choosen as 1.5 or 2 and this procedure can be iterated starting from $D_{i,0}=D_{h/k^i}$ to construct the new estimation
\begin{equation}
 D_{i,j}= \frac{k^{4j}D_{i,j-1}-D_{i-1,j-1}}{k^{4j}-1}
\end{equation}
After $n$ steps, $D_{n,n}$ is of the order of $\mathcal O(h^{4(n+1)})$. Figure \ref{figDer} illustrates in the case of LISA how a relative error of order of $10^{-14}$ on the derivative of $h_{\mu\nu}$ can be reached (in double precision) using Richardson extrapolation. This method does not require to start with a very fine tuned initial step size $h$ and it is possible to stop the iterations when the convergence is sufficient. 

\section{Acknowledgments}

S. Pireaux acknowleges a CNES (Centre National d'Etudes Spatiales, France) post-doctoral grant, plus a one-month contract at the Observatoire de la C\^{o}te d'Azur (OCA, France) as financial support for most part of her work relavent to the present paper. 
A. Hees is research fellow from FRS-FNRS (Belgian Fund for Scientific Research) for his thesis at ORB-UCL (Observatoire Royal de Belgique - Universit\'e Catholique de Louvain, Belgium).


\end{article}
\end{document}